\documentclass[lettersize,journal]{IEEEtran}
\usepackage{amsmath,amsfonts}
\usepackage{algorithmic}
\usepackage{algorithm}
\usepackage{array}
\usepackage[caption=false,font=normalsize,labelfont=sf,textfont=sf]{subfig}
\usepackage{textcomp}
\usepackage{stfloats}
\usepackage{url}
\usepackage{verbatim}
\usepackage{graphicx}
\usepackage{cite}
\begin{document}

\title{VRISE: A Virtual Reality Platfrom for Immersive and Interactive Surveying Education }

\author{Daniel Udekwe, Dimitrios Bolkas, Eren Erman Ozguven, Ren Moses, Qianwen (Vivian) Guo

        % <-this % stops a space

\thanks{Daniel Udekwe, Eren Erman Ozguven, Ren Moses and Qianwen (Vivian) Guo are with the Department of Civil and Environmental Engineering, FAMU-FSU College of Engineering, Tallahassee, Florida, 32310, USA (email: dau24@fsu.edu; eozguven@eng.famu.fsu.edu; moses@eng.famu.fsu.edu; qguo@eng.famu.fsu.edu)}% <-this % stops a space
\thanks{Dimitrios Bolkas is with the Department of Surveying Engineering, The Pennsylvania State University, Wilkes-Barre Campus, Dallas, PA, 18612. (email: dxb80@psu.edu)}
\thanks{Manuscript received xxx, 2025; revised xxx, 2025.}}

%%%\markboth{Journal of \LaTeX\ Class Files,~Vol.~14, No.~8, August~2021}%
%%%{Shell \MakeLowercase{\textit{et al.}}: A Sample Article Using IEEEtran.cls for IEEE Journals}

%%%\IEEEpubid{0000--0000/00\$00.00~\copyright~2021 IEEE}

\maketitle

\begin{abstract}
Surveying is a core component of civil engineering education, requiring students to engage in hands-on spatial measurement, instrumentation handling, and field-based decision-making. However, traditional instruction often poses logistical and cognitive challenges that can hinder accessibility and student engagement. While virtual laboratories have gained traction in engineering education, few are purposefully designed to support flexible, adaptive learning in surveying. To address this gap, we developed Virtual Reality for Immersive and Interactive Surveying Education (VRISE), an immersive virtual reality laboratory that replicates ground-based and aerial surveying tasks through customizable, accessible, and user-friendly modules. VRISE features interactive experiences such as differential leveling with a digital level equipment and waypoint-based drone navigation, enhanced by input smoothing, adaptive interfaces, and real-time feedback to accommodate diverse learning styles. Evaluation across multiple user sessions demonstrated consistent gains in measurement accuracy, task efficiency, and interaction quality, with a clear progression in skill development across the ground-based and aerial surveying modalities. By reducing cognitive load and physical demands, even in tasks requiring fine motor control and spatial reasoning, VRISE demonstrates the potential of immersive, repeatable digital environments to enhance surveying education, broaden participation, and strengthen core competencies in a safe and engaging setting.
\end{abstract}

\begin{IEEEkeywords}
Virtual Reality, Surveying, Differential leveling, Elevation, Virtual lab
\end{IEEEkeywords}

\section{Introduction}
Engineering education relies heavily on hands-on laboratory experiences to foster conceptual understanding, technical competence, and problem-solving skills \cite{bolkas2023collaborative, frady2023use, gomez2022design, may2023online}. In disciplines such as civil engineering, surveying forms a foundational pillar, integrating theoretical knowledge with practical application in spatial measurement, terrain analysis, and geospatial data interpretation \cite{dinis2017virtual, kim2025immersive}. However, traditional surveying instruction,  often dependent on outdoor fieldwork and physical instrumentation, can pose logistical, physical, and educational challenges, including weather dependencies, equipment limitations, and constrained opportunities for repeated practice \cite{qadir2020student, shambare2022critical}.

These longstanding challenges in field-based instruction were further magnified during the COVID-19 pandemic, which brought unprecedented disruptions to global education systems and prompted a rapid shift toward remote and online teaching \cite{adedoyin2023covid}. This transition catalyzed the adoption of more flexible, resilient, and learner-centered instructional models, accelerating the integration of digital technologies in higher education. As a result, online and hybrid learning evolved from emergency responses to strategic solutions for inclusive, accessible, and scalable instruction across disciplines. In fields like civil engineering where experiential learning is critical, this transformation highlighted the urgent need for virtual environments capable of replicating hands-on laboratory experiences. Numerous studies have documented this shift and its broader implications for the future of teaching and learning \cite{bozkurt2020emergency, hodges2020difference, dhawan2020online, daniel2020education, bao2020covid}.

Therefore, virtual engineering laboratories have emerged as a flexible and engaging alternative, offering immersive environments that enhance accessibility, interactivity, and scalability across learning contexts \cite{elmoazen2023learning, luse2021using}. These digital platforms allow students to engage with complex tools and scenarios through simulations and interactive modules that can be accessed on-demand, repeated at the learner’s own pace, and adapted to individual learning styles \cite{alnagrat2022opportunities}. A virtual lab can facilitate their learning by providing multimodal feedback, adjustable pacing, and personalized instruction \cite{javali2025virtual, wahyudi2024understanding}. Grounded in Universal Design for Learning (UDL) principles, virtual labs support diverse pathways for engagement, representation, and expression, making them well-suited to meet the evolving needs of modern engineering classrooms \cite{vetrivel2025impact}.

Despite the growing body of literature on virtual labs, limited research has examined their specific application to surveying education and even fewer studies have explored their feasibility for students \cite{damianova2025serious}. Surveying, traditionally taught through field-based training with instruments like theodolites, total stations, and GPS receivers, requires spatial visualization and motor coordination, skills that may pose barriers for students with cognitive or physical differences \cite{bolkas2022first}. A virtual laboratory designed to simulate surveying tasks can mitigate these challenges by allowing repeated practice, instant feedback, and contextual visualization in a controlled and distraction-free digital environment \cite{deriba2024assessment, prasetya2023utilizing}.

Moreover, recent advances in 3D visualization, game-based learning, and virtual reality (VR) platforms enhance the realism and interactivity of virtual surveying environments \cite{vzilak2022systematic}. Such technologies can simulate real-world terrain, instruments, and measurement tasks, helping students grasp abstract surveying concepts while building confidence and technical proficiency \cite{rezaei2024applications}. Integrating assistive technologies, such customizable user interfaces, and cognitive support tools, further augments the accessibility and efficacy of virtual labs for students \cite{ntoa2024digital, sable2020analysis}.

In response to the limitations of traditional field-based surveying education, this paper presents VRISE, a novel virtual reality platform that simulates essential surveying tasks in an immersive and accessible digital environment. Unlike conventional VR implementations that rely on generic control schemes or static simulations, VRISE offers highly interactive, task-specific modules for differential leveling and drone-based waypoint navigation, replicating both the procedural and cognitive demands of real-world surveying. A distinctive technical feature of VRISE is its integration of a single exponential smoothing (SES) algorithm, which dynamically stabilizes controller inputs to reduce jitter and enhance the precision of virtual instrument manipulation. This implementation enables more natural and accurate user interaction, addressing a common barrier in VR-based skill training. Through its combination of pedagogically grounded design, technical innovation, and performance-driven evaluation, VRISE represents a unique and scalable approach to transforming how surveying is taught in engineering education.

The remainder of the paper is organized as follows: Section \ref{sec:lit-rev} reviews existing literature on virtual laboratories, surveying education, and inclusive learning technologies. Section \ref{sec:method} describes the materials and methods used in developing VRISE. Section \ref{results} presents preliminary results and system evaluations. Finally, Section \ref{conclusion} summarizes the conclusions and outlines current limitations and future directions for this work.

\section{LITERATURE REVIEW} \label{sec:lit-rev}
Engineering education is undergoing a fundamental transformation, driven by the rapid advancement of immersive technologies such as Virtual Reality (VR), Augmented Reality (AR), Mixed Reality (MR), and game-based learning systems. These innovations are reshaping how complex engineering concepts are taught, practiced, and assessed, especially in domains that require spatial reasoning, procedural accuracy, and technical skill, such as surveying. This literature review explores two interrelated areas of research: first, the evolution of surveying in the digital era, (Section 2.1); and second, the applications of VR across various branches of engineering education (Section 2.2), emphasizing how immersive environments can enhance access, learning outcomes, and student engagement. By synthesizing existing findings, this review identifies the educational opportunities associated with deploying virtual laboratories and immersive simulations, thereby providing a foundation for the development and evaluation of VRISE as an equity-centered VR learning platform.

\subsection{The Evolution of Surveying in the Digital Age}
 The surveying profession has seen major advancements in recent decades, driven by digital technologies such as small Unmanned Aircraft Systems. (sUASs), Light Detection and Ranging (LiDAR), and Global Navigation Satellite Systems (GNSS). These tools have fundamentally altered how spatial data is collected, processed, and interpreted. Table \ref{tab:comp1} outlines the key differences between traditional and digital surveying practices, highlighting how modern technologies have introduced new skills, expanded professional responsibilities, and improved accessibility in surveying education. The modern surveyor has evolved from data collectors to data managers and information providers.

 \begin{table*}[!t]
\centering
\caption{Comparison of Traditional and Digital-Age Surveying Practices} \label{tab:comp1}
\begin{tabular}{p{3cm}p{4.5cm}p{5.5cm}p{2cm}}

\hline
\textbf{Dimension} & \textbf{Traditional Surveying} & \textbf{Digital-Age Surveying} & \textbf{References} \\
\hline
Tools \& Instrumentation & Theodolites, total stations, manual tapes & UAVs, LiDAR scanners, GNSS, total stations with digital output & \cite{bolkas2022first, rezaei2024applications, lu2023traditional} \\
\\
Data Collection \& Storage & Paper notebooks, manual sketches & Real-time digital recording, cloud synchronization, automated backups & \cite{elmoazen2023learning, waters2021teaching} \\
\\
Processing Techniques & Manual calculations, visual estimations & Point cloud processing, automated stitching, GIS analysis & \cite{hernandez2020engineering, vzilak2022systematic, pavelka2024using} \\
\\
Required Competencies & Field setup, measuring, reading analog instruments & 3D modeling, scripting, spatial analytics & \cite{dinis2017virtual, ntoa2024digital} \\
\\
Role of Surveyor & Field technician, measurement executor & Geospatial data scientist, technical consultant in multidisciplinary projects & \cite{gomez2022design, yeganeh2025future} \\
\\
Measurement Accuracy & Dependent on physical skill, subject to human error & Dependent on instrument specifications and measurement methods & \cite{rezaei2024applications, abd2023virtual} \\
\\
Accessibility & Physical presence required, inaccessible to students with disabilities & UDL, VR customizations, cognitive aids & \cite{elfakki2023efficient, ntoa2024digital, sable2020analysis} \\
\\
Learning Outcomes & Procedural competency, limited repeatability & Conceptual mastery, repeatable and self-paced learning & \cite{lampropoulos2024virtual, wahyudi2024understanding, yeganeh2025future}\\
\\
Collaboration Style & Isolated field crews, hierarchical reporting & Interdisciplinary coordination using shared digital environments & \cite{javali2025virtual, vetrivel2025impact} \\
\\
Assessment Tools & Instructor’s observations, static rubrics & Analytics dashboards, interaction logs, adaptive feedback systems & \cite{elmoazen2023learning, prasetya2023utilizing} \\
\hline
\end{tabular}
\end{table*}

\subsection{VR Applications in Engineering Education}
VR is becoming an integral part of engineering education due to its ability to simulate real-world scenarios, provide hands-on practice, and enhance engagement \cite{wang2018critical}. It addresses limitations of traditional methods such as cost, accessibility, and safety. Table \ref{tab:comp2} illustrates how VR is being adopted across engineering disciplines, showing its benefits and domain-specific use cases.

\begin{table*}[!t]
\centering
\caption{VR Applications Across Engineering Domains} \label{tab:comp2}
\begin{tabular}{p{3.1cm}p{4.2cm}p{4cm}p{4.8cm}}
\hline
\textbf{Domain} & \textbf{Traditional Approach} & \textbf{VR Enhancement} & \textbf{Key Benefit and References} \\
\hline
Architecture and Design & Static blueprints, CAD renderings & VR walkthrough of building and urban designs & Improved spatial perception, design validation \newline \cite{ochoa2016virtual, sampaiovirtual}\\
\\
Mechanical Engineering & Prototyping with physical parts and CNC tools & Simulation of assemblies and parts in VR & Hands-on learning, safe prototyping \newline \cite{gavish2015evaluating, pletz2020evaluation}  \\
\\
Chemical Engineering & Lab-based distillation, titration and process equipment & Simulations of chemical reactions, process control, and fluid dynamics in hazardous environment & Safety, repeatability, remote access \newline \cite{alhalabi2016virtual, jensen2018review}  \\
\\
Structural Engineering & Manual calculations and models for load distribution & Interactive modeling and load simulation in VR environment & Enhanced conceptual clarity, immersive experience of structural failures \newline \cite{henstrom2023immersive, wang2018critical} \\
\\
Environmental Engineering & GIS maps, environmental sampling kits & Virtual ecosystem for pollution tracking, water testing, and spatial modeling & Engagement, visualization of abstract data, simulated interventions \newline \cite{larsson2023visualizing, try2021virtual, zhu2025virtual} \\
\\
Biomedical Engineering & Cadavers, physical mannequins & VR anatomical labs, simulated surgeries and interactive physiology modules & Safe exploration, ethical use, enhanced visual-spatial learning \newline \cite{moro2017effectiveness, venkatesan2021virtual} \\
\\
Civil Engineering & Physical structural models, lab-based stress and load testing & Virtual stress testing, bridge building, terrain analysis & Cost reduction, scalability, real-time feedback \newline \cite{dinis2017virtual, rezaei2024applications, rudolfa2024inclusive} \\
\\
Surveying & Outdoor fieldwork using GPS, total station, and leveling rods & Immersive simulations of total station setup, sUAS surveying, and differential leveling & Accessibility, repeatability, reduced weather dependency \newline \cite{bolkas2022first, mcduff2025enhancing, vzilak2022systematic} \\
\hline
\end{tabular}
\end{table*}

\subsection{Summary and Synthesis}
The adoption of virtual laboratories in engineering education has gained substantial momentum in recent years, offering scalable, safe, and flexible alternatives to traditional hands-on learning environments. This transition is particularly impactful in scenarios where physical constraints, such as limited equipment, safety risks, or geographic barriers, restrict access to conventional labs. Research has shown that virtual labs can successfully replicate complex engineering procedures, promote analytical thinking, and enhance technical proficiency through iterative practice and real-time feedback mechanisms \cite{frady2023use, hassan2022virtual}

Despite this progress, the specific application of virtual laboratory platforms in surveying education remains relatively limited. Traditional surveying instruction often depends on manual manipulation of specialized instruments such as total stations and GPS receivers, tasks that require precise motor control, strong spatial reasoning, and on-the-fly decision-making \cite{bolkas2022first, waters2021teaching}. These demands can be particularly challenging for some students, who may struggle with executive function, working memory, and spatial processing \cite{elfakki2023efficient, hernandez2020engineering}.

Emerging technologies such as Virtual Reality (VR), Augmented Reality (AR), Mixed Reality (MR) and game-based simulations offer promising alternatives by providing immersive, controlled environments in which students can repeatedly engage with key surveying tasks. These platforms allow users to visualize terrain, operate instruments, and interpret geospatial data with minimal distraction and physical strain, features especially beneficial for learners with neurodiverse profiles \cite{yeganeh2025future, vzilak2022systematic}

Importantly, VR-based systems are increasingly being designed with accessibility in mind, incorporating features such as text-to-speech narration, adaptive user interfaces, and embedded cognitive supports. Guided by the principles of UDL, these technologies aim to provide multiple pathways for engagement, expression, and comprehension, thus broadening participation among students with diverse educational needs \cite{ntoa2024digital, sable2020analysis}

Nevertheless, key challenges remain. Studies have identified gaps between the capabilities of existing virtual lab platforms and the specific instructional requirements of civil engineering curricula, particularly in supporting students who need tailored cognitive support \cite{wahyudi2024understanding, yenduri2023assistive}. Addressing this gap, the VRISE initiative represents a forward-looking effort to create an inclusive, VR-based surveying platform \cite{christian2021virtual}. By embedding assistive technologies and adaptive feedback systems into immersive learning modules, VRISE seeks to ensure equitable access to core civil engineering competencies while supporting the full spectrum of learner diversity.

\section{MATERIALS AND METHODS} \label{sec:method}
An overview is presented in this section of the development platform, hardware specifications, and user interface components that underpin the system’s functionality.

\subsection{VRISE Technical Framework}
This section presents the technical framework of the VRISE system, encompassing its hardware and software configuration. It also details the implementation of velocity smoothing algorithms designed to enhance input stability and ensure a seamless user experience within the immersive environment.

\subsubsection{Hardware and Software Specifications}
The primary hardware employed is the Meta Quest 3 headset, selected for its wireless capabilities, high-resolution displays (2064 × 2208 pixels per eye), and integrated inside-out tracking. The Meta Quest 3 model features a Snapdragon XR2 Gen 2 processor, 8GB of RAM, and support for Wi-Fi 6E, ensuring smooth, responsive interactions critical for reducing cognitive load and motion sickness among users. The device’s hand tracking, adjustable refresh rates (up to 120Hz), and ergonomic controllers enhance accessibility and user experience, offering multiple modalities of interaction suited to diverse learning preferences.

The virtual environments were developed using Unity 2022 LTS, utilizing the Universal Render Pipeline (URP) for optimal graphical performance while maintaining lightweight processing demands. Unity’s XR Plugin Management and OpenXR integration were implemented to enable seamless compatibility with the Meta Quest platform. Accessibility features were embedded through assets like the XR Interaction Toolkit and audio-visual augmentation plugins, while performance enhancements such as Level of Detail (LOD) systems, baked lighting, and occlusion culling, applied to maintain high frame rates and minimize latency.
A Dell Precision workstation powered the development and testing processes, featuring an Intel Core i7-14700k, 3.40GHz processor, NVIDIA RTX 4090 graphics card, 64GB of DDR5 RAM, and a 2TB NVMe SSD. This high-performance system enables efficient asset baking, scene compilation, and high-fidelity simulation testing without interruptions, ensuring a robust pipeline from design to deployment.

To guarantee optimal user experience, the VR Subspace build settings were finely tuned. Key configurations included:
\begin{enumerate}
    \item Utilization of OpenGLES3 as the sole Graphics API to ensure compatibility and stability.
    \item Adoption of a Linear color space to improve lighting realism.
    \item Anti-aliasing set to 2x or 4x MSAA for visual clarity without significant performance degradation.
    \item Application of baked lighting and reflection probes, eliminating real-time global illumination demands.
    \item Frame rate targeting of 72–90 FPS, vital for minimizing motion sickness.
    \item Texture compression standardized by ASTC to optimize asset memory footprint.
\end{enumerate}

\subsubsection{Velocity Smoothening and Deadzone Stabilization}
In the development of VRISE's VR Subspace, ensuring fluid, predictable, and comfortable user interaction was paramount, particularly given the accessibility requirements of students new to surveying. Early testing revealed that the raw positional and velocity data from the Meta Quest controllers exhibited high-frequency noise and small-amplitude hand tremors, which resulted in jittery object interactions. 

To address this, a Single Exponential Smoothing (SES) approach was selected. This method, widely recognized for its simplicity and effectiveness in real-time signal stabilization, is adopted to recursively filter out noise from the series of hand motions \cite{udekwe2025human} 

Mathematically, the smoothing is applied follows:

\begin{equation}
    p_{sm} = \alpha \cdot p_r(t) + (1-\alpha) \cdot p_{sm} (t-1)
\end{equation}
\begin{equation}
    v_{sm}(t) = \alpha \times v_r(t) + (1-\alpha) \times v_{sm}(t-1)
\end{equation}

where $p_{sm} (t)$  and $v_{sm} (t)$ are the current smoothed position and velocity values, $p_r(t)$  and $v_r(t)$ are the raw input position and velocity values, and $\alpha $ is the smoothing factor, set experimentally at 0.2 for typical users.

By substituting the prior smoothed terms, these equations can be expanded and expressed in summation form as \cite{udekwe2025virtual}:

\begin{equation}
        v_{sm}(t) = \alpha \sum_{i=1}^{t-2}(1-\alpha)^{i-1} \cdot v_r(t-i) + (1-\alpha)^{t-2} \cdot v_{sm} (t-1)
\end{equation}

\begin{equation}
    p_{sm}(t) = \alpha \sum_{i=1}^{t-2} (1-\alpha)^{i-1} \cdot p_r (t-i)+(1-\alpha)^{t-2} \cdot p_{sm}(t-1)
\end{equation}

\begin{figure}[!t]
  \centering
  \includegraphics[width=0.5\textwidth]{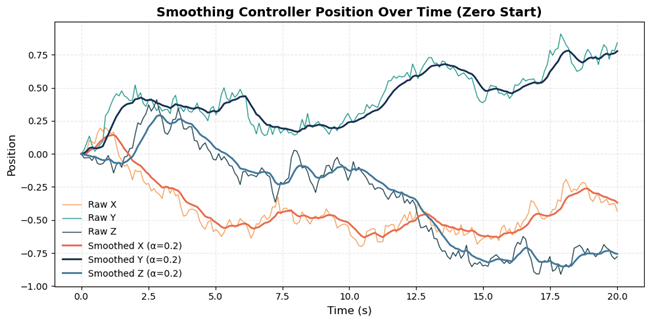}
      \caption{Controller Position Smoothing on X, Y, Z Axes (SES $\alpha $ = 0.2)} \label{fig:cont}
\end{figure}

\begin{figure}[!t]
  \centering
  \includegraphics[width=0.5\textwidth]{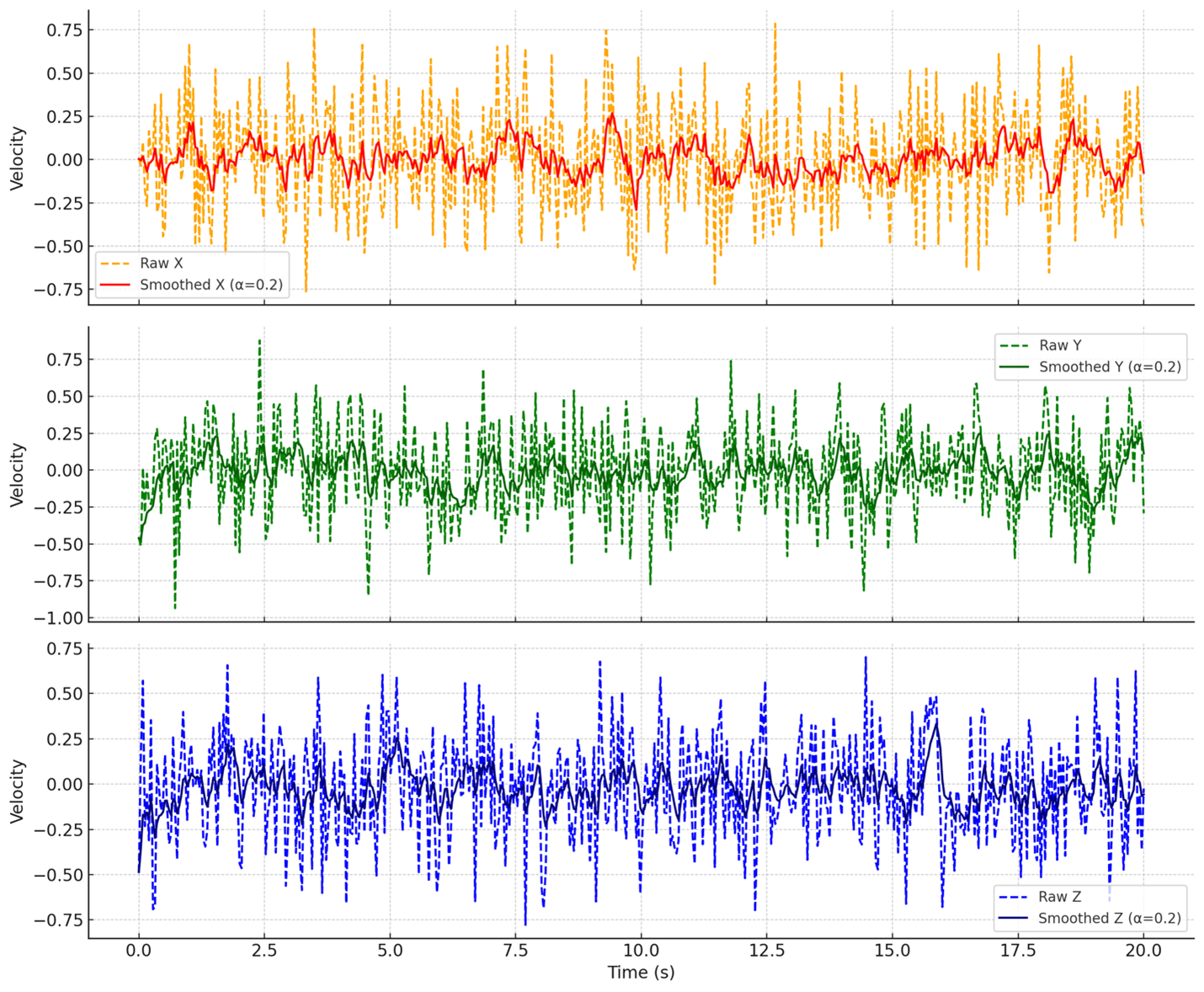}
      \caption{Controller Velocity Smoothing on X, Y, Z Axes (SES $\alpha $ = 0.2)} \label{fig:cont2}
\end{figure}

Performance analysis is conducted by logging raw and smoothed position and velocity values during a standard object manipulation task involving picking up, moving, and rotating objects within the VR environment. The logged data demonstrates a marked reduction in high-frequency noise post-smoothing while preserving major motion trends essential for user feedback. Figure \ref{fig:cont} and Figure \ref{fig:cont2} illustrate the comparison between the raw and smoothed positional and velocity profiles captured during a 20-second manipulation sequence.

Furthermore, to enhance interaction stability after velocity smoothing, a dead zone was introduced to eliminate small, unintended micro-movements caused by residual noise from VR controllers. If the smoothed velocity falls below a threshold, it is treated as zero, effectively filtering out minor tremors and preventing visual jitter. This ensures only deliberate movements affect virtual objects, improving control, comfort, and accessibility especially for users with fine motor challenges.

\subsection{Experiment Design}
In this section, the components of the experiment are presented, including the orientation scene, the SFS, and the interactive tablet interface. 

\subsubsection{Orientation Scene}
\begin{figure*}[!t]
\centering
\subfloat[]{\includegraphics[width=0.455\textwidth]{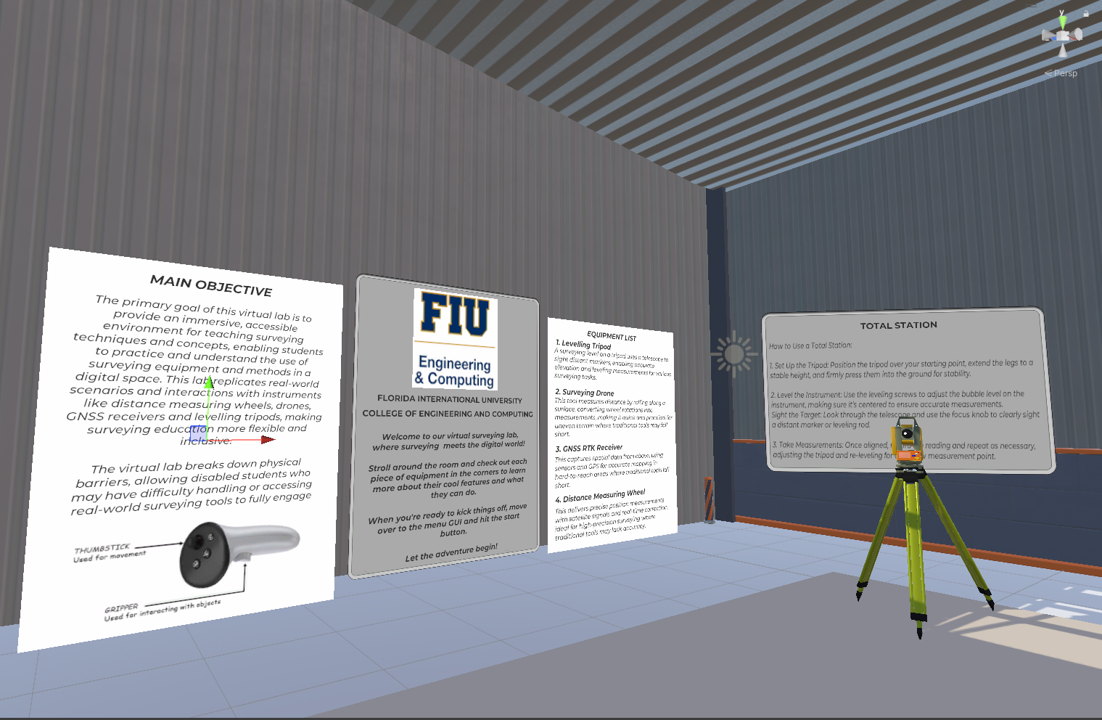}%
\label{}}
\hfil
\subfloat[]{\includegraphics[width=0.48\textwidth]{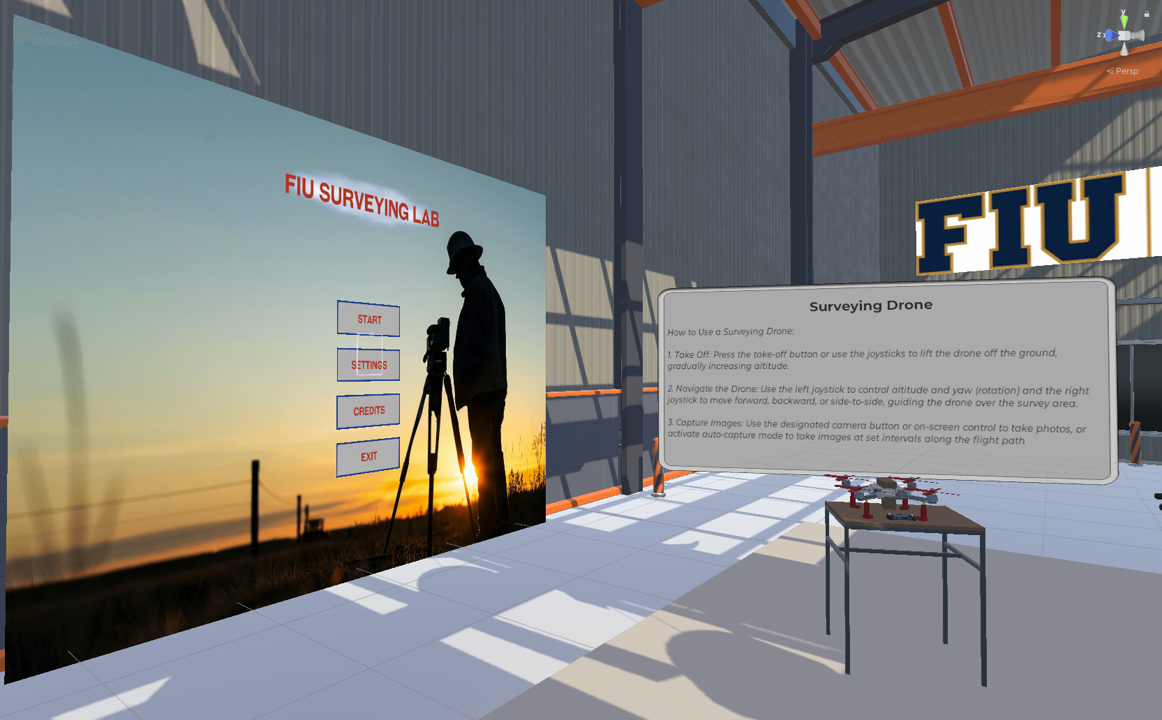}%
\label{}}
\caption{Orientation Scene Showing the Instruction Panel and Interactive Display Walls: (a) Instructional Panel Wall; (b) Interactive Display Wall}
\label{fig:orientation_scene}
\end{figure*}

The Orientation Scene in Figure \ref{fig:orientation_scene} serves as the introductory environment within the VR Subspace VRISE, designed to familiarize users with the virtual lab, communicate objectives, and prepare students for interactive tasks. Set in a visually realistic lab, it includes a large display panel with key information such as the purpose, procedures, and controls for operating surveying sUASs and instruments. The streamlined interface offers four main options, namely, Start, Settings, About, and Exit, allowing users to begin the simulation, adjust accessibility settings, learn about the lab context, or exit the application. 

A key design consideration for the Orientation Scene was cognitive simplicity: by minimizing visual clutter, using large clear fonts, and providing concise, instructional text, the environment reduces cognitive load and supports students with diverse learning needs. Additionally, the scene’s realistic architectural setting helps bridge the gap between virtual and real-world civil engineering labs, fostering a greater sense of immersion and relevance.

\subsubsection{Surveying Field Scene (SFS)}

\begin{figure*}[!t]
\centering
\subfloat[]{\includegraphics[width=0.48\textwidth]{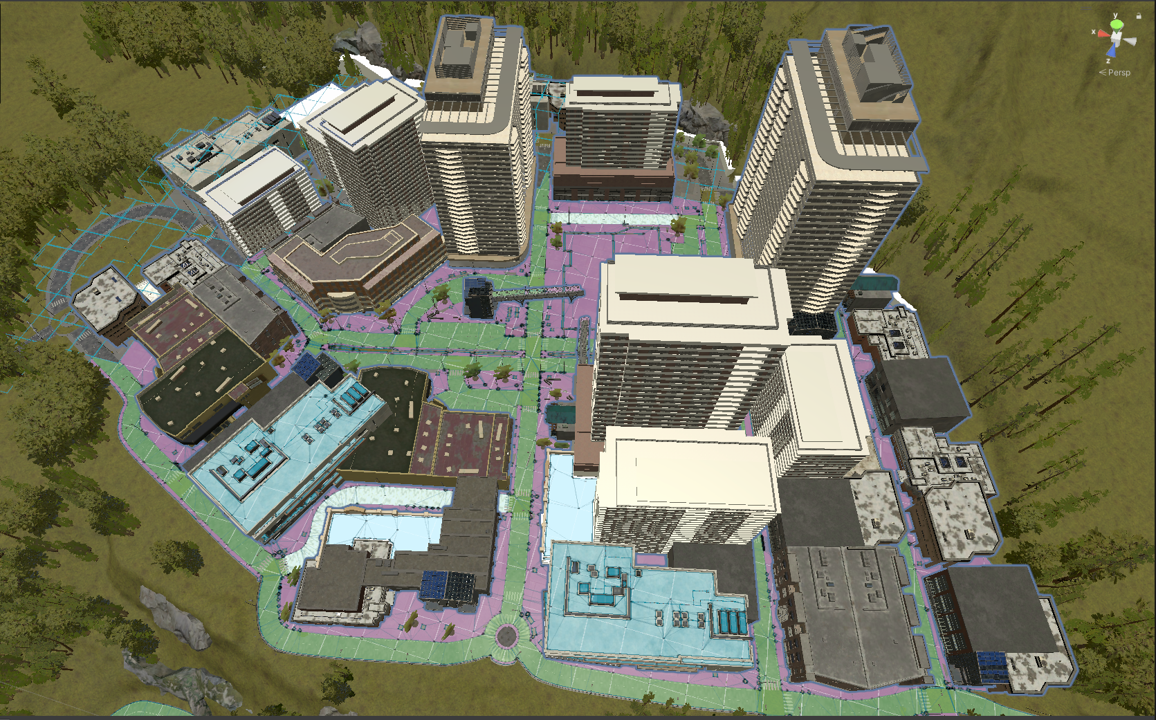}\label{fig:panel_wall}}
\hfil
\subfloat[]{\includegraphics[width=0.48\textwidth]{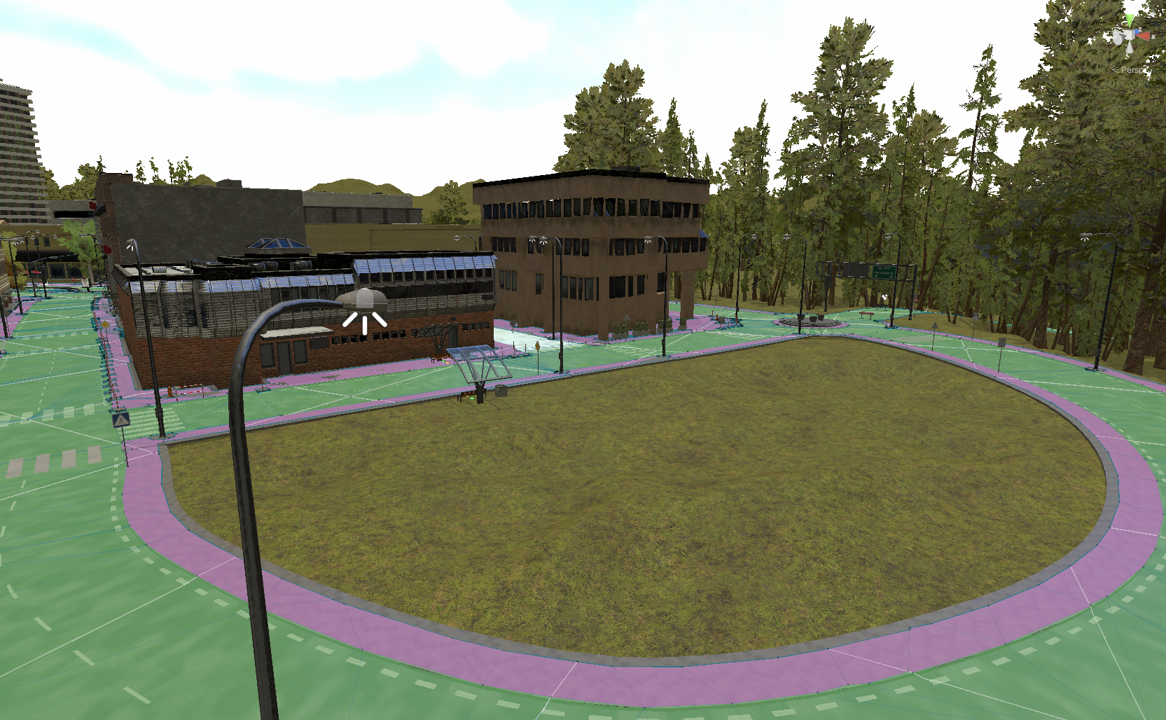}\label{fig:display_wall1}}
\hfil
\subfloat[]{\includegraphics[width=0.48\textwidth]{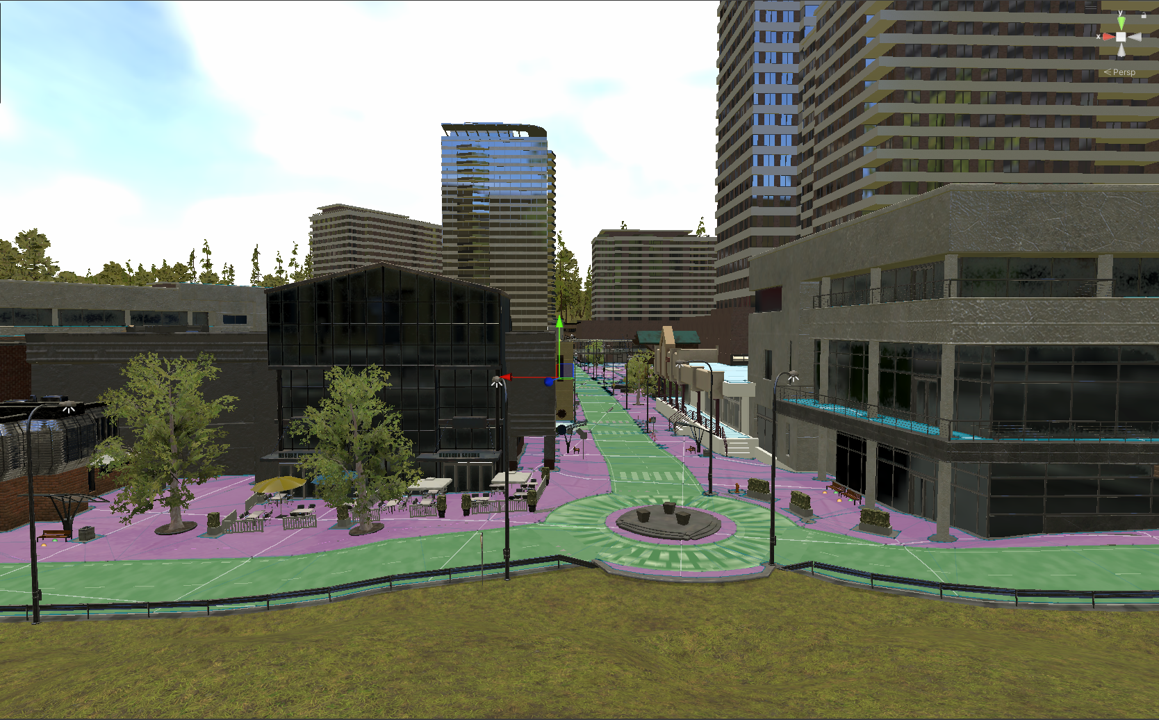}\label{fig:display_wall2}}
\hfil
\subfloat[]{\includegraphics[width=0.48\textwidth]{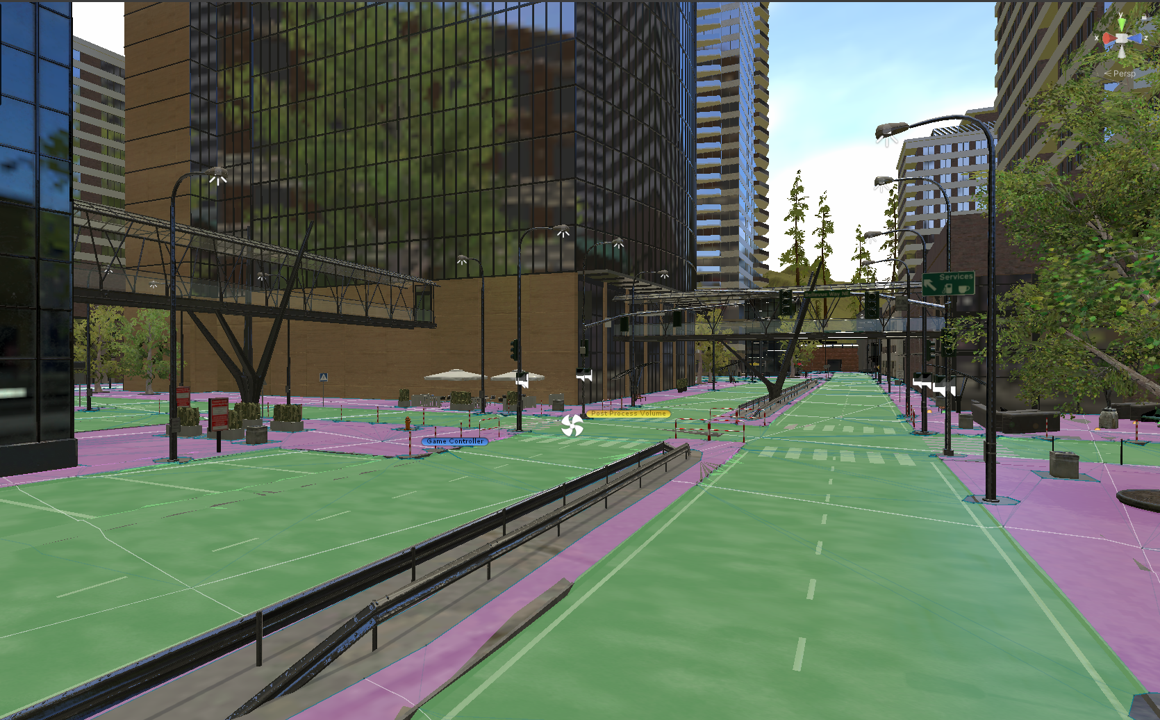}\label{fig:display_wall3}}
\caption{SFS Showing Several Scenes: (a) Aerial Overview; (b) Open Parkland Area; (c) Urban Corridor; (d) Street Level Perspective}
\label{fig:sfs}
\end{figure*}

The SFS is the main operational environment within the VRISE's VR Subspace, designed to replicate complex outdoor conditions for surveying tasks as shown in Figure \ref{fig:sfs}. Set in a detailed virtual urban district featuring high-rise buildings, institutional facilities, parklands, pathways, and roads, it offers a dynamic landscape for practicing equipment use and techniques. The scene simulates real-world surveying challenges such as structural occlusions, elevation changes, variable sightlines, and diverse surface materials, encouraging students to adapt their strategies across both open and constrained environments, an essential skill for civil engineering and surveying practice.

A critical aspect of the SFS’ design is its scale: the environment is spatially expansive, encouraging meaningful navigation, planning, and measurement, rather than limiting users to static tasks. Realistic daylighting is also simulated, introducing variability into surveying accuracy and visibility, thereby enriching the educational experience.

\subsubsection{Interactive Table Interface}
\begin{figure*}[!t]
\centering
\subfloat[]{\includegraphics[width=0.48\textwidth]{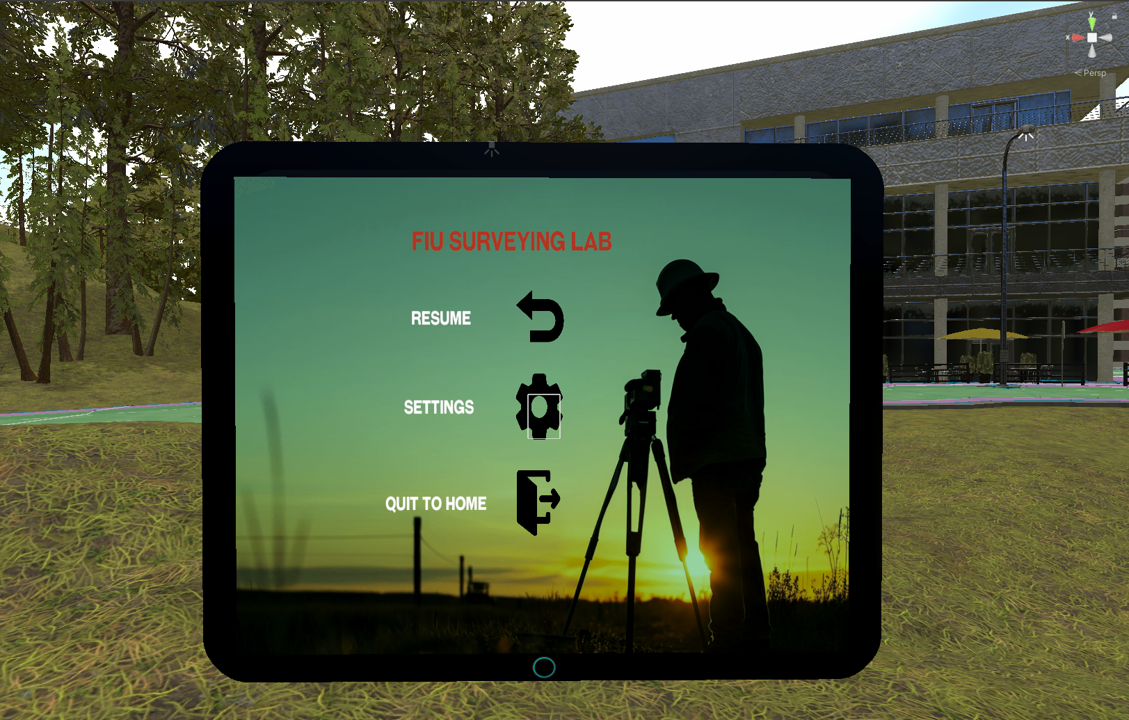}%
\label{}}
\hfil
\subfloat[]{\includegraphics[width=0.48\textwidth]{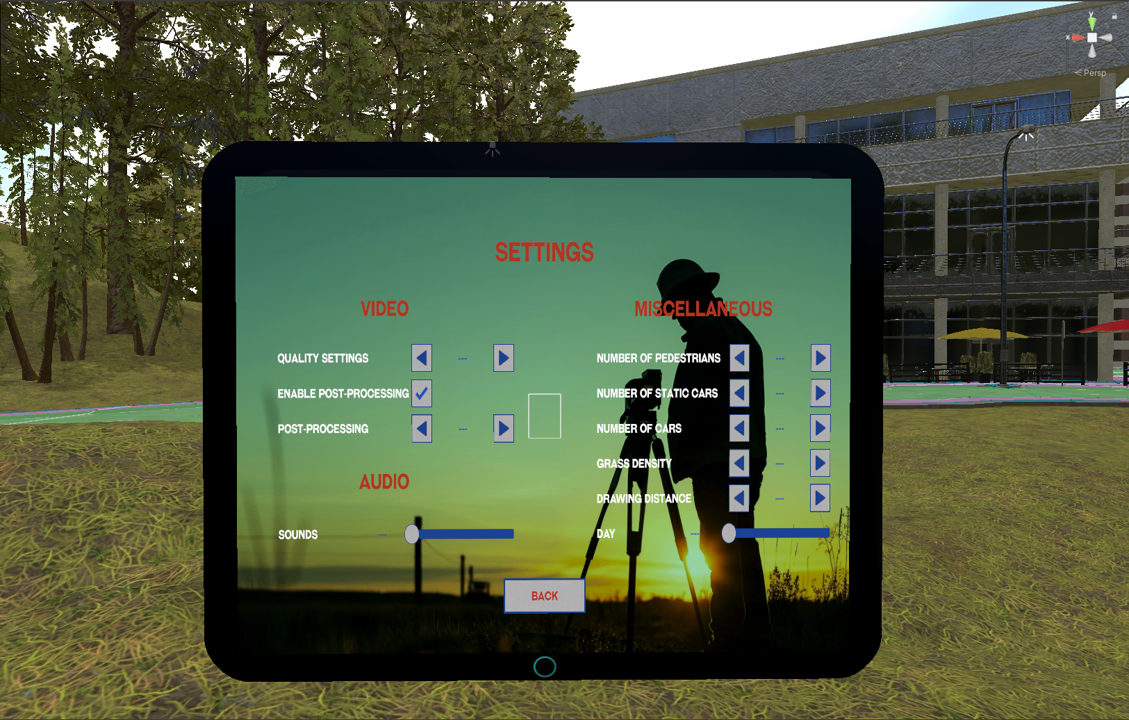}%
\label{}}
\caption{Interactive Tablet Interface Showing the Main Menu and Settings Menu Screens: (a) Main Menu Screen; (b) Setting Menu Screen }
\label{fig:tablet_display}
\end{figure*}

An interactive tablet interface within the SFS allows users to manage settings and session controls in real time as shown in Figure \ref{fig:tablet_display}. Featuring large, high-contrast icons and immediate feedback, the menu enables users to resume, adjust settings, or exit the simulation. 

Within the settings menu, users can adjust video quality, toggle post-processing effects, modify audio levels, and configure miscellaneous parameters such as grass density and the visibility of motorists as well as pedestrians. This level of customization allows students to optimize the VR experience according to their personal preferences, device performance, or educational needs. By placing control directly in the user's hands without disrupting the immersive experience.

\subsection{Surveying Tasks Modules}
This section describes the ground-based and aerial surveying techniques employed in the study, using leveling equipment and sUAS, respectively.

\subsubsection{Ground Based Surveying}
\begin{figure*}[!t]
\centering
\subfloat[]{\includegraphics[width=0.486\textwidth]{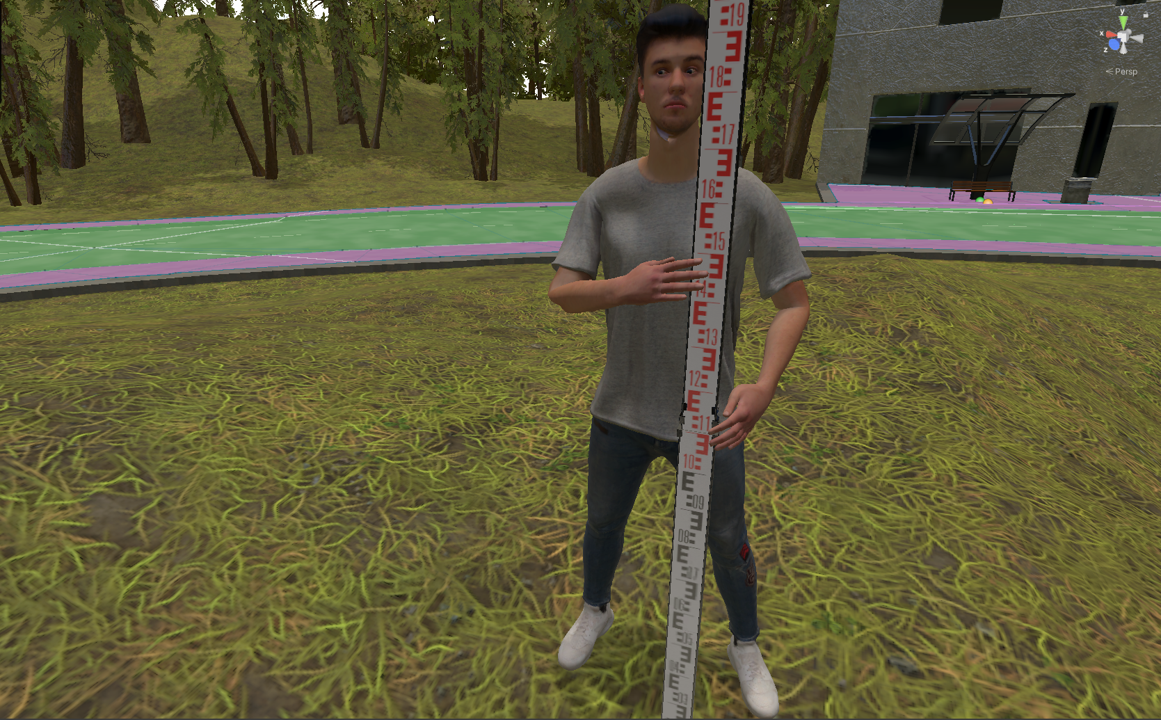}\label{fig:main_menu}}
\hfil
\subfloat[]{\includegraphics[width=0.48\textwidth]{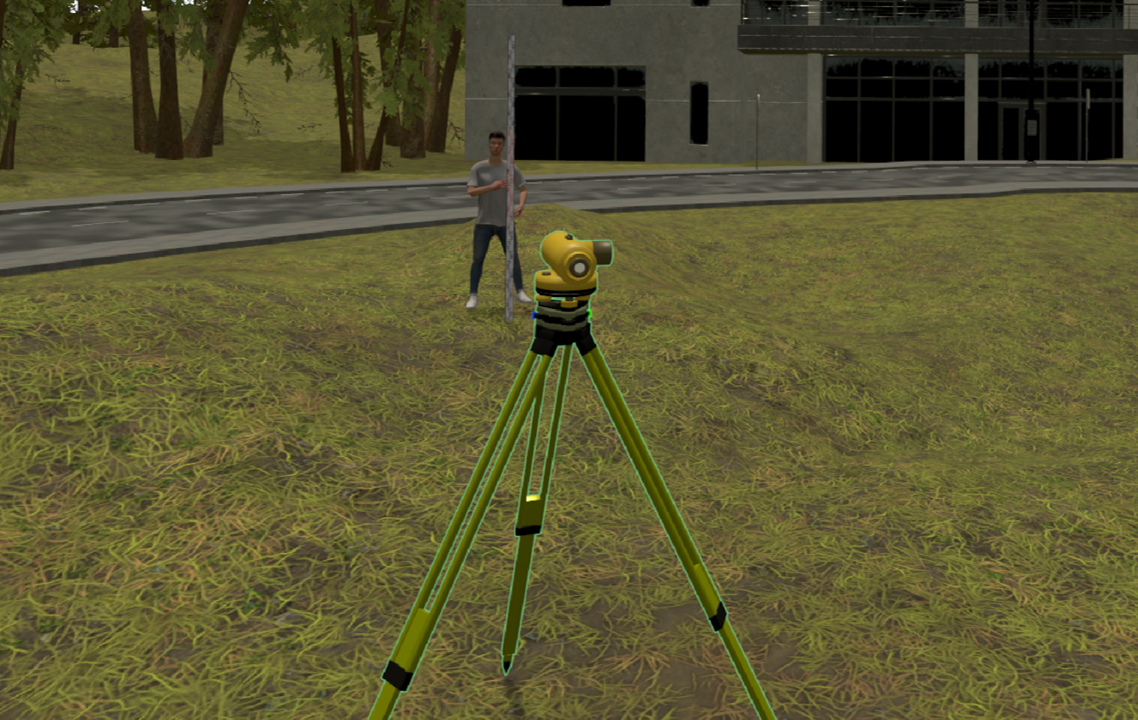}\label{fig:settings_menu}}
\caption{Components of ground-based surveying: (a) Leveling rod with surveying; (b) Digital level mounted on tripod set}
\label{fig:ground_surveying}
\end{figure*}

Ground-based leveling exercises within the SFS simulate core surveying procedures involving the use of a leveling rod and a digital level. Students collaborate with a virtual surveying assistant who positions the leveling rod at designated points, while they operate a digital level mounted on a tripod to record elevation data as shown in Figure \ref{fig:ground_surveying}. The task workflow includes initial tripod setup, instrument alignment, bubble leveling, and target acquisition, providing a highly realistic replication of standard surveying techniques.

\subsubsection{Tripod Positioning}
\begin{figure*}[!t]
\centering
\subfloat[]{\includegraphics[width=0.47\textwidth]{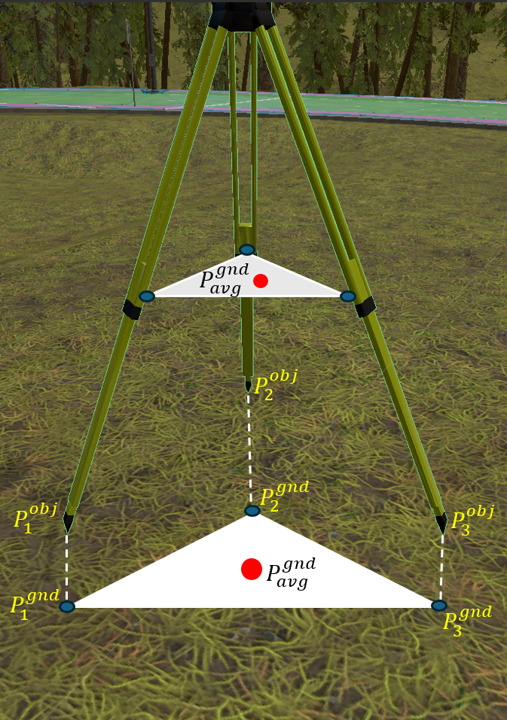}\label{}}
\hfil
\subfloat[]{\includegraphics[width=0.47\textwidth]{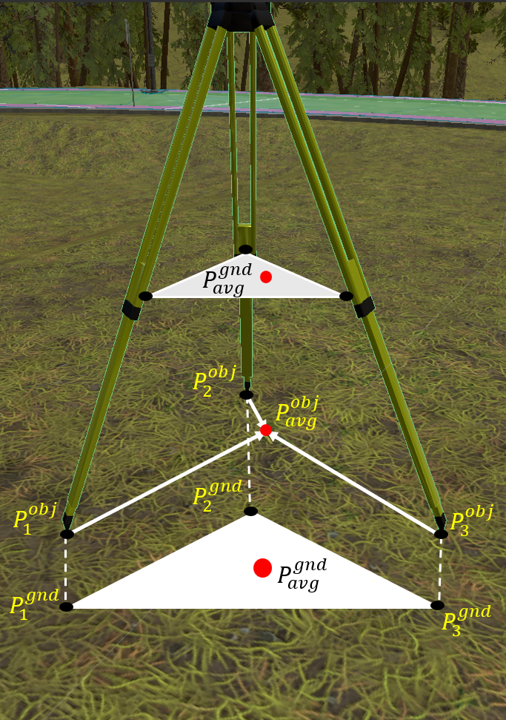}\label{}}
\caption{Tripod positioning: (a) Vertical Measurement Setup; (b) Object Position Estimation}
\label{fig:tripod_positioning}
\end{figure*}

Accurate tripod positioning is essential for ensuring the stability and alignment of digital level measurements in any ground-based surveying activity. In the VR SFS, a precise method is implemented to simulate the tripod’s placement using geometric averaging of contact points as illustrated in Figure \ref{fig:tripod_positioning}. The system considers three physical ground contact points of the tripod, represented as \( P_1^{\text{gnd}}, P_2^{\text{gnd}}, P_3^{\text{gnd}} \), and their corresponding object-space projections \( P_1^{\text{obj}}, P_2^{\text{obj}}, P_3^{\text{obj}} \), which define the virtual tripod leg contact geometry. The tripod positioning is implemented similarly to the approach presented in \cite{bolkas2021surveying}.

To evaluate the tripod’s alignment, the average positions of both the object-level (tripod model) and the terrain-level (virtual ground) triangle vertices are computed. These are derived as the centroids of the respective triangles using the following equations:

\begin{equation}
\mathbf{P}_{\text{avg}}^{\text{obj}} = \left[ \frac{1}{n} \sum_{i=1}^{n} x_i^{\text{obj}}, \; \frac{1}{n} \sum_{i=1}^{n} y_i^{\text{obj}}, \; \frac{1}{n} \sum_{i=1}^{n} z_i^{\text{obj}} \right]
\end{equation}

\begin{equation}
\mathbf{P}_{\text{avg}}^{\text{gnd}} = \left[ \frac{1}{n} \sum_{i=1}^{n} x_i^{\text{gnd}}, \; \frac{1}{n} \sum_{i=1}^{n} y_i^{\text{gnd}}, \; \frac{1}{n} \sum_{i=1}^{n} z_i^{\text{gnd}} \right]
\end{equation}

where \( n = 3 \) corresponds to the tripod's three legs. This averaging provides a stable and consistent reference for aligning the digital level above the ground centroid. The closer the vertical projection of \( \mathbf{P}_{\text{avg}}^{\text{obj}} \) aligns with \( \mathbf{P}_{\text{avg}}^{\text{gnd}} \), the more accurately the tripod is centered and leveled over the ground station, minimizing alignment error in subsequent surveying tasks.

After computing the centroids \( \mathbf{P}_{\text{avg}}^{\text{obj}} \) and \( \mathbf{P}_{\text{avg}}^{\text{gnd}} \), the distance between them gives a direct measure of how misaligned the tripod is over the survey point:

\begin{equation}
d = \sqrt{(\mathbf{P}_1^{\text{obj}} - \mathbf{P}_1^{\text{gnd}})^2 + (\mathbf{P}_2^{\text{obj}} - \mathbf{P}_2^{\text{gnd}})^2 + (\mathbf{P}_3^{\text{obj}} - \mathbf{P}_3^{\text{gnd}})^2}
\end{equation}

This equation quantifies how far off the digital level is from being properly centered over the ground station. Ideally, \( d \rightarrow 0 \) for correct setup.

If the legs of the tripod do not lie on a flat surface, the triangle they form may not be planar, which can lead to wobble. You can evaluate coplanarity using the normal vector cross product. Let the three ground points be \( \vec{A}, \vec{B}, \vec{C} \). Then compute:

\begin{equation}
\vec{n} = (\vec{B} - \vec{A}) \times (\vec{C} - \vec{A})
\end{equation}

To assess planarity, check whether the magnitude of \( \vec{n} \) is stable (i.e., not close to zero), and optionally, evaluate the height variance among \( z_i \) values:

\begin{equation}
\sigma_z = \sqrt{ \frac{1}{n} \sum_{i=1}^{n} (z_i - \bar{z})^2 }
\end{equation}

\begin{figure*}[!t]
\centering
\subfloat[]{\includegraphics[width=0.497\textwidth]{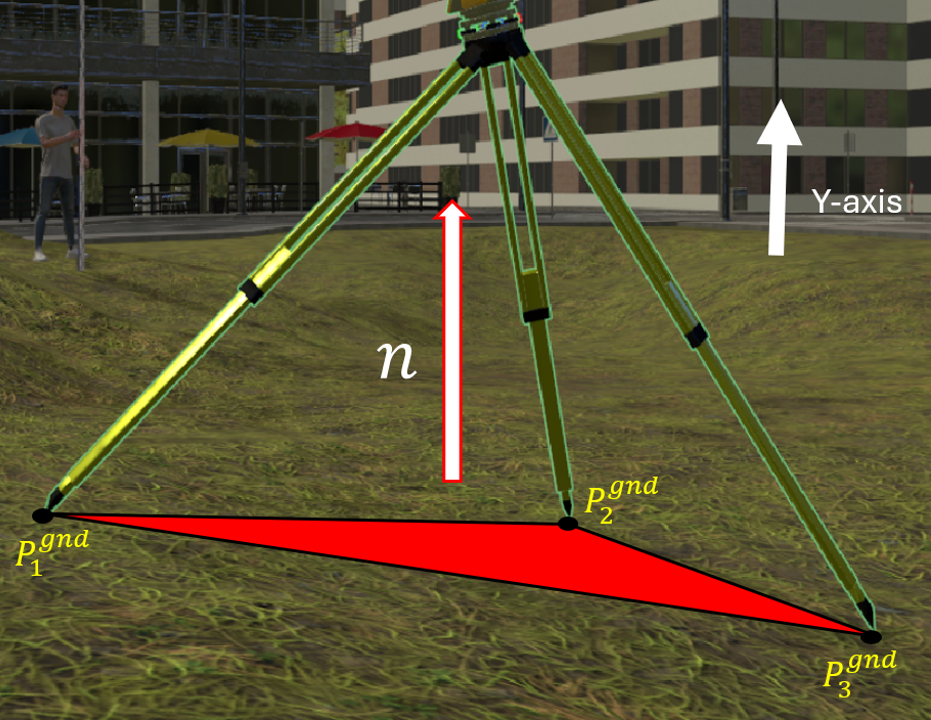}\label{}}
\hfil
\subfloat[]{\includegraphics[width=0.45\textwidth]{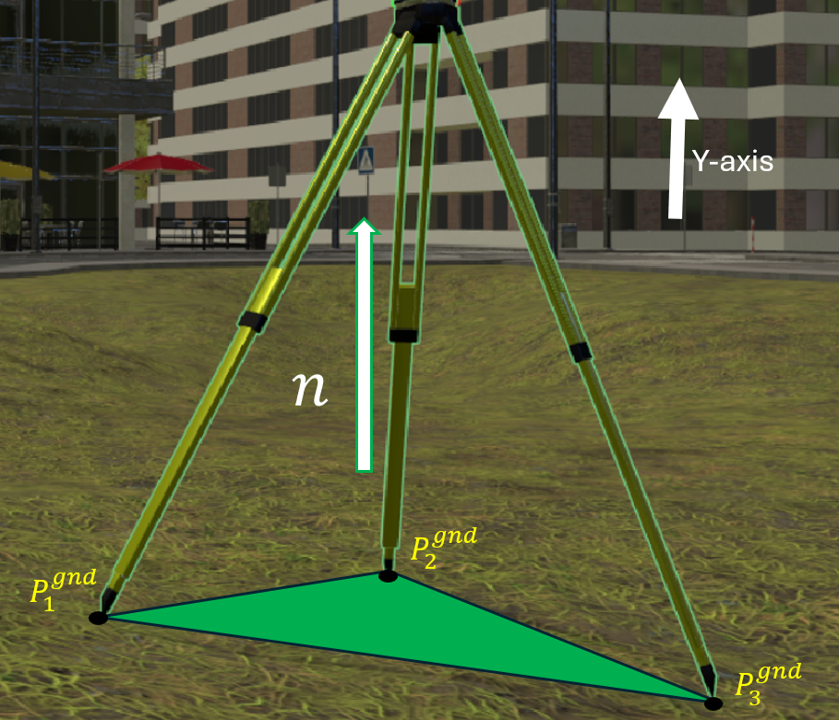}\label{}}
\caption{Ground Plane Normal Estimation for Alignment of Digital Level: (a) Tilted Ground Plane; (b) Leveled Ground Plane}
\label{fig:tripod_leveling}
\end{figure*}

This helps to ensure that the tripod is not tilted unevenly due to the terrain slope as illustrated in Figure \ref{fig:tripod_leveling}.

\subsubsection{Tripod Levelling}
This section describes the two-step leveling method used to align the digital level. The initial phase involves coarse leveling by adjusting the tripod legs, while the second phase focuses on fine leveling using the tribrach screws to achieve precise instrument alignment. This procedure is consistent with the approach detailed in \cite{bolkas2021surveying}

\begin{figure*}[!t]
\centering
\subfloat[]{\includegraphics[width=0.47\textwidth]{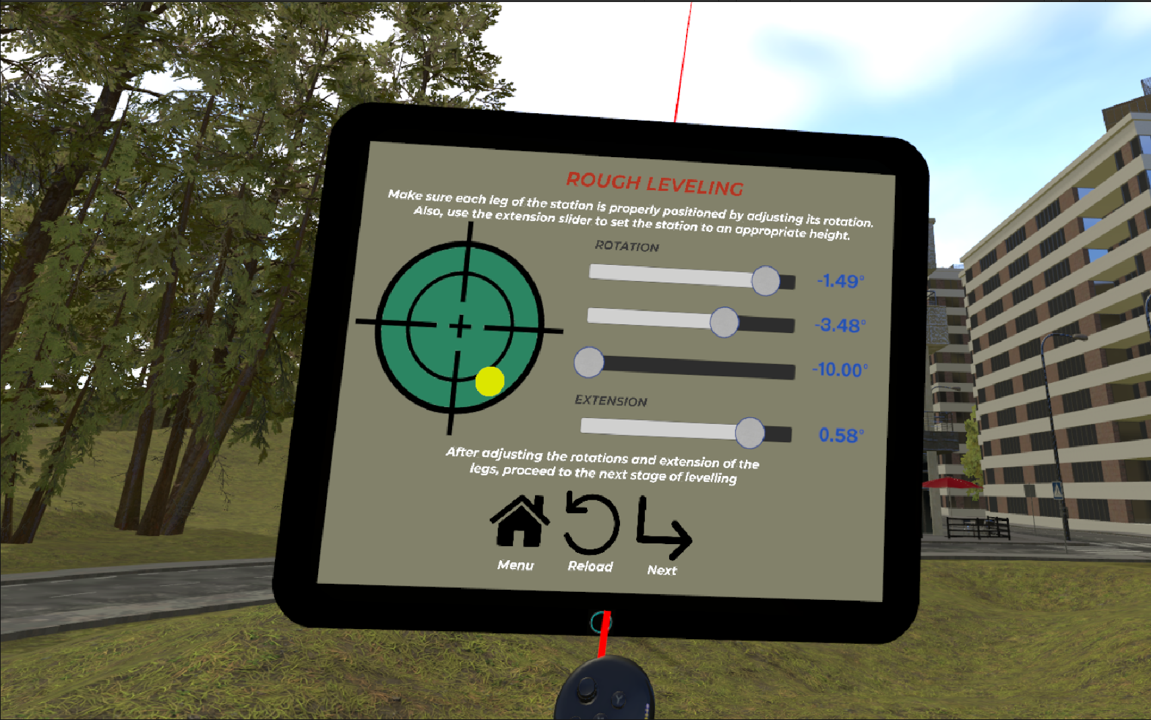}\label{}}
\hfil
\subfloat[]{\includegraphics[width=0.47\textwidth]{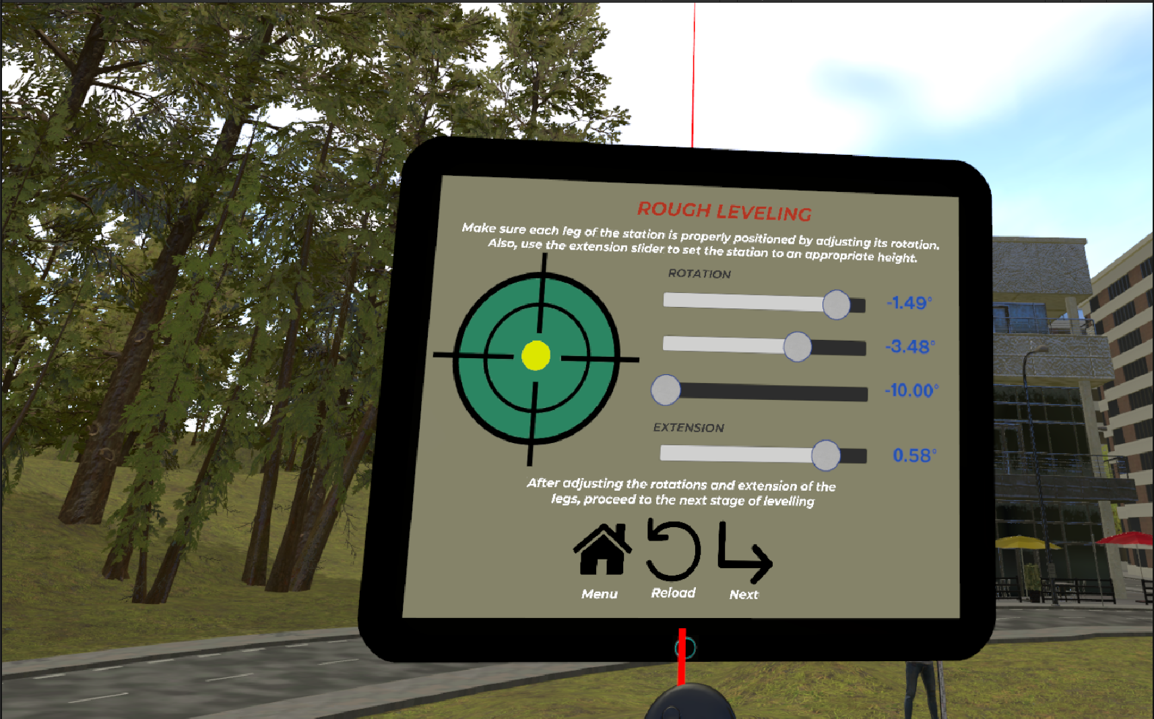}\label{}}
\caption{Rough leveling interface: (a) Initial state showing misalignment; (b) Final State after adjustments}
\label{fig:rough_leveling}
\end{figure*}

Rough Levelling:\\
Rough leveling is the preliminary step in tripod and digital level setup, ensuring that the instrument is approximately aligned and stable before engaging in fine precision adjustments. In the VRISE virtual environment, this process is supported by an interactive interface that allows users to manipulate the virtual tripod by adjusting its rotational orientation and leg extension length via on-screen sliders as illustrated in Figure \ref{fig:rough_leveling}. The interface also visualizes leveling accuracy using a simulated circular level, enabling learners to receive immediate visual feedback during the process.

Mathematically, the leveling correction is modeled using an angular displacement approximation defined by two compound variables: \( u \) and \( v \), representing the total horizontal and vertical misalignment of the tripod, respectively. These are derived from combined rotational offsets:

\begin{equation}
u = u' + u'', \quad v = v' + v'',
\end{equation}

where \( u' \) and \( v' \) represent the base plate tilt due to terrain, and \( u'' \), \( v'' \) represent user-corrected adjustments. The instrument’s horizontal and vertical alignment is then approximated using the following second-order corrected equations:

\begin{equation}
X = u \sqrt{1 - \frac{v^2}{2}}
\end{equation}

\begin{equation}
Z = v \sqrt{1 - \frac{u^2}{2}}
\end{equation}

These equations ensure that combined tilt adjustments preserve orthogonality, reflecting real-world leveling physics, where pitch and roll corrections affect each other non-linearly at higher angles. As users manipulate the interface to minimize these offsets, the system visually confirms leveling convergence through the centering of the bubble in the target reticle. Once the tilt parameters fall within acceptable margins, users can proceed to the precise leveling phase.

To quantify the bubble offset (as observed by the user in the circular level), the following expressions are used:

\begin{equation}
\theta_x = \arctan\left( \frac{\Delta x}{r} \right)
\end{equation}

\begin{equation}
\theta_y = \arctan\left( \frac{\Delta y}{r} \right)
\end{equation}

where \( \Delta x \) and \( \Delta y \) are the pixel or unit offsets of the bubble from the center of the level, \( r \) is the calibrated radius of the level circle, and \( \theta_x \), \( \theta_y \) represent the angular deviation in the X and Y axes (i.e., leveling error).

\begin{figure*}[!t]
\centering
\subfloat[]{\includegraphics[width=0.47\textwidth]{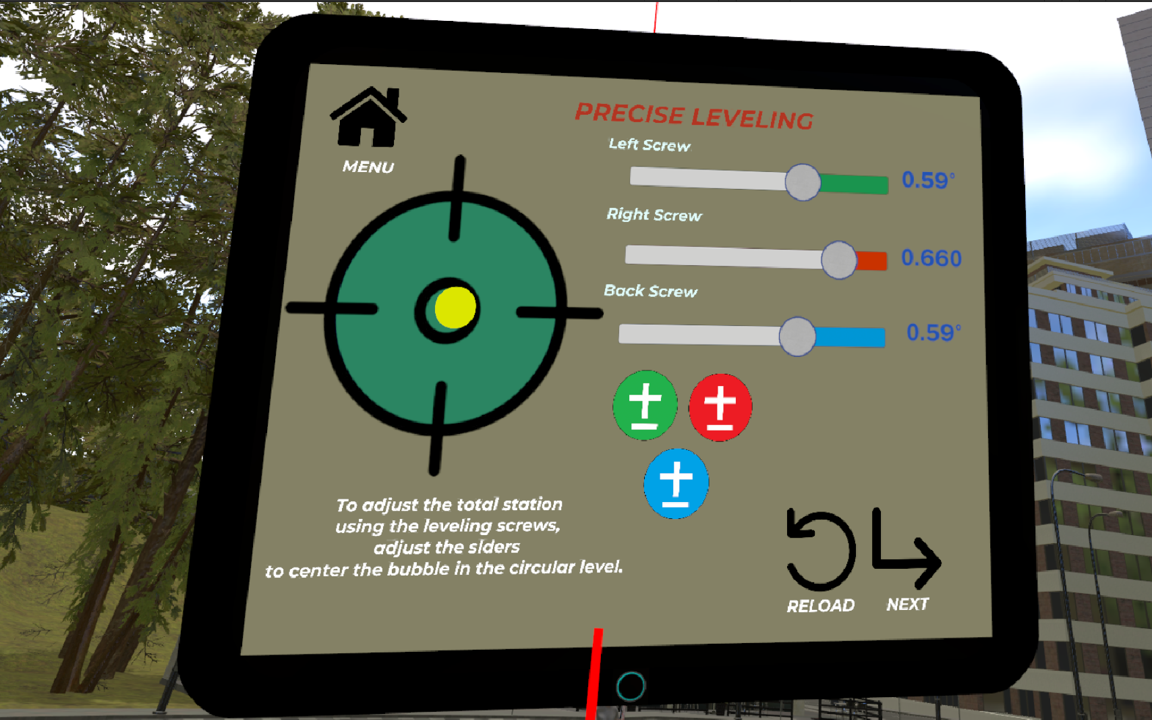}\label{}}
\hfil
\subfloat[]{\includegraphics[width=0.47\textwidth]{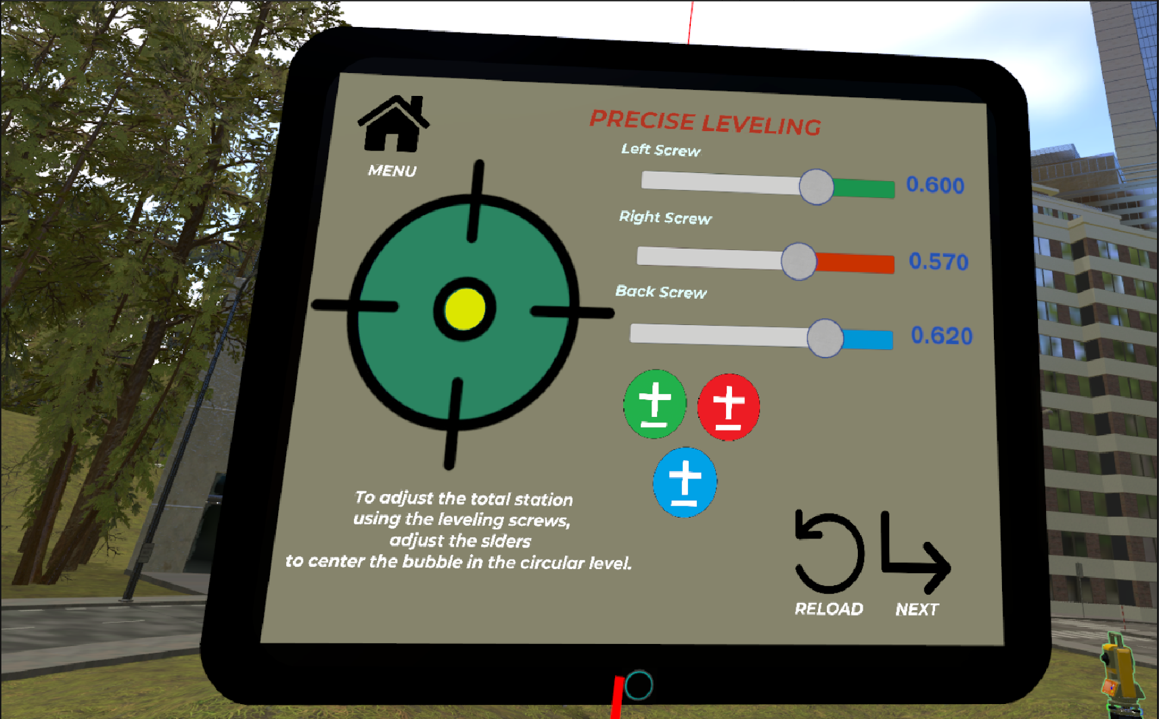}\label{}}
\caption{Precise Leveling Interface using the tribarch screws: (a) Initial State with bubble offset; (b) Final state after screw adjustment }
\label{fig:precise_leveling}
\end{figure*}

Precise Leveling:\\
Precise leveling is the final calibration step performed after rough leveling, enabling users to align the digital level’s vertical axis perfectly perpendicular to the local horizontal plane. In the VRISE environment, this process is simulated through an interactive circular level interface where users adjust three tribrach screws left, right, and back to fine-tune the orientation of the instrument as illustrated in Figure \ref{fig:precise_leveling}. The objective is to center the bubble within the circular level by applying precise rotational offsets using screw controls. 

The leveling adjustments are modeled using screw-driven displacement equations that quantify how each rotation alters the instrument's horizontal orientation. The adjustments along two axes, \( u'' \) and \( v'' \), are derived from screw input values:

\begin{equation}
u'' = \frac{l}{2} - \frac{r}{2}
\end{equation}

\begin{equation}
v'' = b - \left( \frac{l}{2} + \frac{r}{2} \right)
\end{equation}

where \( l \), \( r \), and \( b \) represent the rotation displacements of the left, right, and back screws respectively, and \( u'' \), \( v'' \) are the fine horizontal corrections in the X and Y axes.

As users make adjustments, the circular bubble's position responds in real time, reflecting the net effect of \( u'' \) and \( v'' \) in moving the instrument toward vertical alignment. The system encourages iterative tuning, using color-coded indicators to promote accuracy. Once the bubble is fully centered and screw values stabilize, users can proceed to data collection, confident that the leveling error is within the accepted tolerance.

If each screw rotation causes a known angular change \( \alpha \), then the incremental corrections are given by:

\begin{equation}
\delta u'' = \alpha \cdot \frac{\Delta l - \Delta r}{2}
\end{equation}

\begin{equation}
\delta v'' = \alpha \cdot \left( \Delta b - \frac{\Delta l + \Delta r}{2} \right)
\end{equation}

where \( \Delta l \), \( \Delta r \), and \( \Delta b \) represent the actual changes in screw values (e.g., per click or per degree), and \( \alpha \) is a conversion factor such as degrees per millimeter of screw movement.

\subsubsection{Aerial Surveying using sUASs}
\begin{figure*}[!t]
\centering
\subfloat[]{\includegraphics[width=0.47\textwidth]{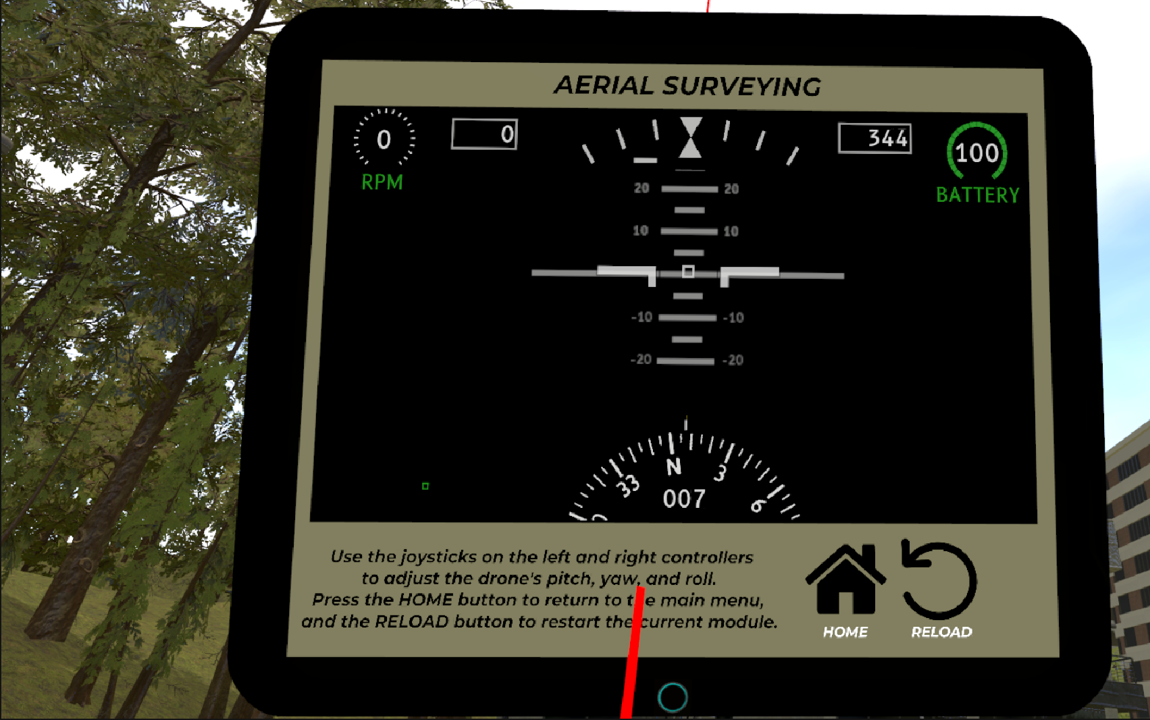}\label{fig:drone_control}}
\hfil
\subfloat[]{\includegraphics[width=0.44\textwidth]{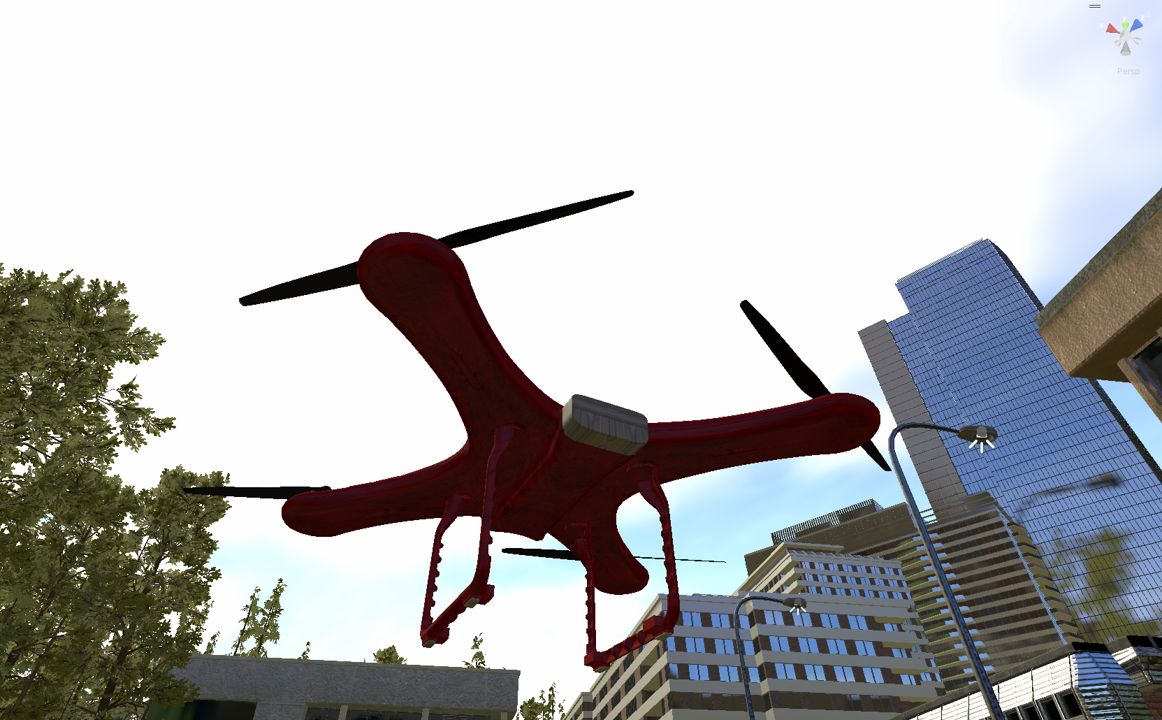}\label{fig:drone_flight}}
\caption{Screenshots from the aerial surveying module showing (a) the drone control interface and (b) the virtual drone in flight.}
\label{fig:aerial_surveying_module}
\end{figure*}

The aerial surveying module in VRISE provides students with an immersive experience in sUAS-based topographic data collection, integrating both flight control mechanics and visual instrumentation as shown in Figure \ref{fig:aerial_surveying_module}. Through the developed tablet interface in VR, students are introduced to a realistic heads-up display (HUD) that includes telemetry data such as rotor revolutions per minute (RPM), battery level, pitch, roll, yaw indicators, and directional heading. Users operate the virtual drone using controller joysticks to manipulate the drone’s orientation and trajectory in three-dimensional space.

\subsection{Task Setup and Feasibility Planning}
This section provides a description of the differential leveling and waypoint trailing exercises using the digital level and sUAS  respectively. 

\subsubsection{Differential Leveling Exercise}
\begin{figure}[!t]
  \centering
  \includegraphics[width=0.5\textwidth]{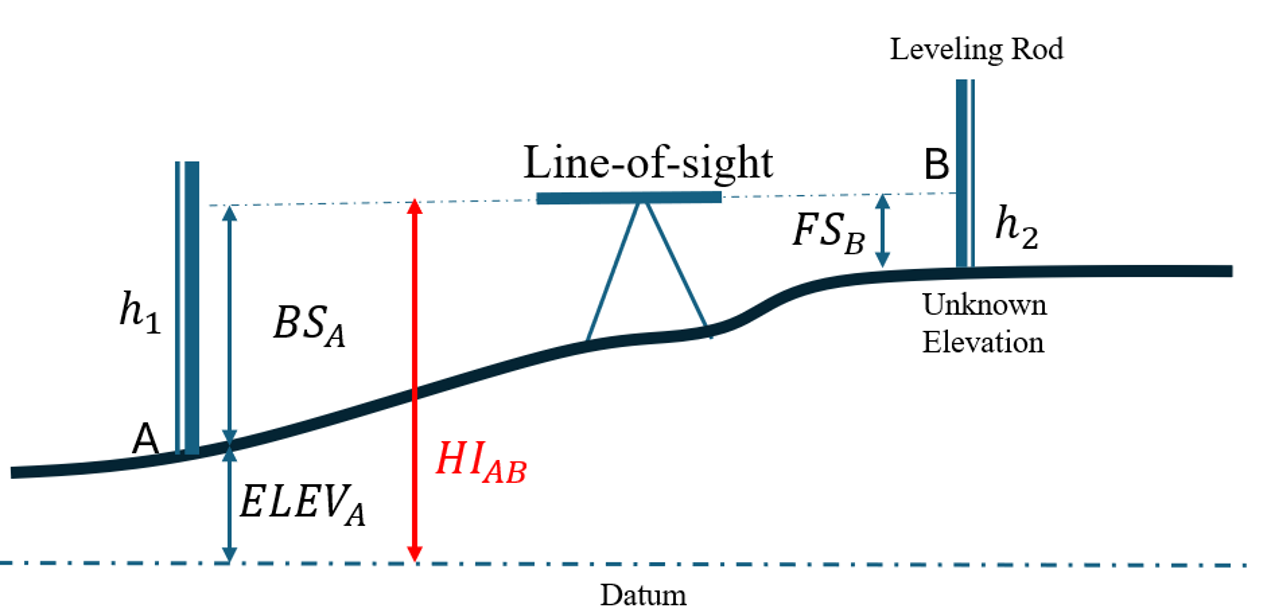}
      \caption{Conceptual Diagram of Differential Leveling} \label{fig:diff_leveling_concept}
\end{figure}

In this exercise, students determine the elevation of an unknown point B using the known elevation at point A and measurements taken through a digital level. As shown in Figure \ref{fig:diff_leveling_concept}, the digital level is set up between two points, one with a known elevation $(ELEV_A)$ and one with an unknown elevation $(ELEV_B)$. 

The student first takes a back sight (\( \text{BS}_A \)) reading on a leveling rod placed at point A, which is used to compute the height of instrument (HI):

\begin{equation}
\text{HI}_{AB} = \text{ELEV}_A + \text{BS}_A
\end{equation}

\begin{figure}[!t]
  \centering
  \includegraphics[width=0.5\textwidth]{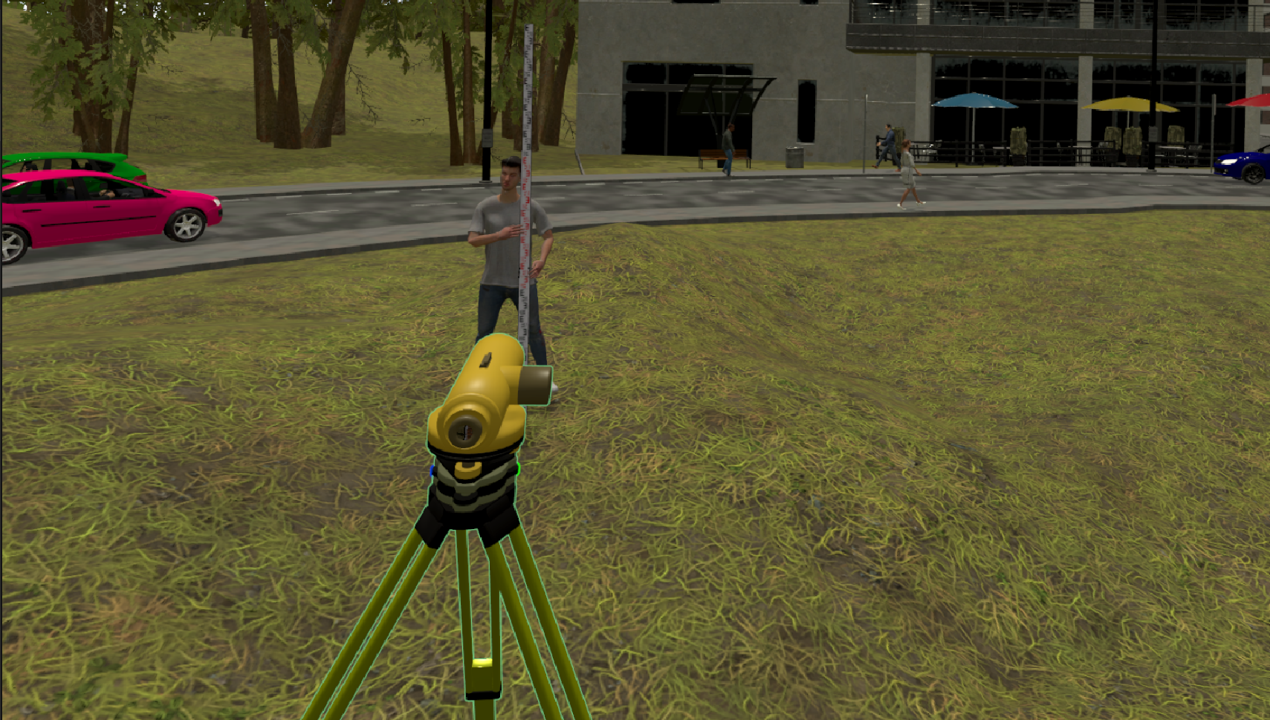}
      \caption{Setup of the differential leveling exercise} \label{fig:leveling_example}
\end{figure}

Next, a foresight (\( \text{FS}_B \)) reading is taken at point B, as illustrated in Figure~\ref{fig:leveling_example}. The elevation at point B is then calculated using:

\begin{equation}
\text{ELEV}_B = \text{HI}_{AB} - \text{FS}_B
\end{equation}

This hands-on VR task teaches students how to apply core leveling equations in the field by simulating real-world workflows. It reinforces their understanding of line-of-sight principles and height transfer using surveying instruments.

\subsubsection{Waypoint trailing exercise}
\begin{figure}[!t]
  \centering
  \includegraphics[width=0.5\textwidth]{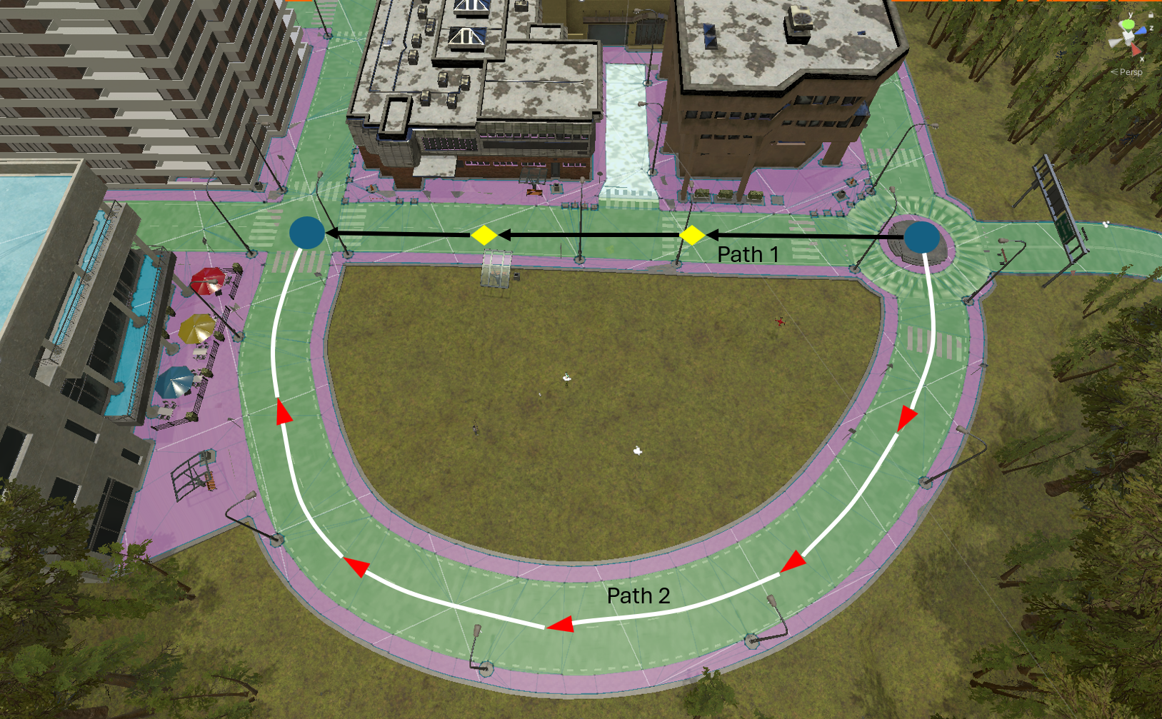}
      \caption{Waypoint Trailing Exercise with Independent Paths} \label{fig:waypoint_trailing}
\end{figure}

The waypoint trailing exercise is designed to train users in autonomous drone navigation and spatial awareness within the virtual surveying environment. It comprises two independent flight paths,  Path 1 and Path 2, each consisting of a series of sequential waypoints that the drone must follow in a predefined order as shown in Figure \ref{fig:waypoint_trailing}.

Path 1 is a straight-line route intended to introduce users to basic forward motion and heading control. It helps users develop a feel for drone responsiveness and accuracy in maintaining linear trajectories. In contrast, Path 2 features a curved trajectory that challenges users to perform continuous heading adjustments and lateral movements, simulating more complex real-world navigation scenarios.

\subsection{Evaluation Metrics}
To evaluate student performance within the virtual surveying environment of VRISE, a series of quantitative and qualitative metrics were developed. These metrics assess not only the technical accuracy of completed tasks but also the precision, efficiency, and level of support required during each activity. The goal is to holistically measure skill acquisition, cognitive engagement, and task execution fidelity in comparison to conventional field training.

\begin{itemize}
    \item \textbf{Positional Accuracy:}  
    For leveling and tripod placement tasks, the system calculates the deviation between the expected and actual computed elevations or alignments.  
    For example, in the differential leveling exercise, the error in the final computed elevation at point B is given by:
    \begin{equation}
    \epsilon_{\text{elev}} = \left| \frac{h_B^{\text{comp}} - h_B^{\text{act}}}{h_B^{\text{act}}} \right|
    \end{equation}
    where \( \epsilon_{\text{elev}} \) is the elevation error, \( h_B^{\text{comp}} \) is the computed elevation at point B, and \( h_B^{\text{act}} \) is the actual elevation.

    \item \textbf{Task Completion Time:}  
    The time taken to complete each subtask (e.g., tripod positioning, leveling, measurement, drone flight path completion) is logged. This metric captures procedural fluency and familiarity with task sequences.

    \item \textbf{Interaction Efficiency:}  
    The number of input attempts (e.g., screw rotations, slider adjustments, or joystick corrections) before convergence is tracked.  
    A high number of adjustments may indicate user uncertainty, while fewer, more accurate inputs reflect procedural mastery.
\end{itemize}

\section{RESULTS AND DISCUSSION} \label{results}
This section presents the preliminary findings from both ground-based and aerial-based surveying exercises. The experiments were conducted by a single participant using VR, who possessed only a basic level of familiarity with both VR technology and surveying techniques. These initial results offer insights into user performance and interaction patterns under minimal prior training, helping to establish a baseline for future comparisons with more experienced users or refined system configurations

\subsection{GROUND SURVEYING}
\begin{figure*}[!t]
\centering
\subfloat[]{\includegraphics[width=0.47\textwidth]{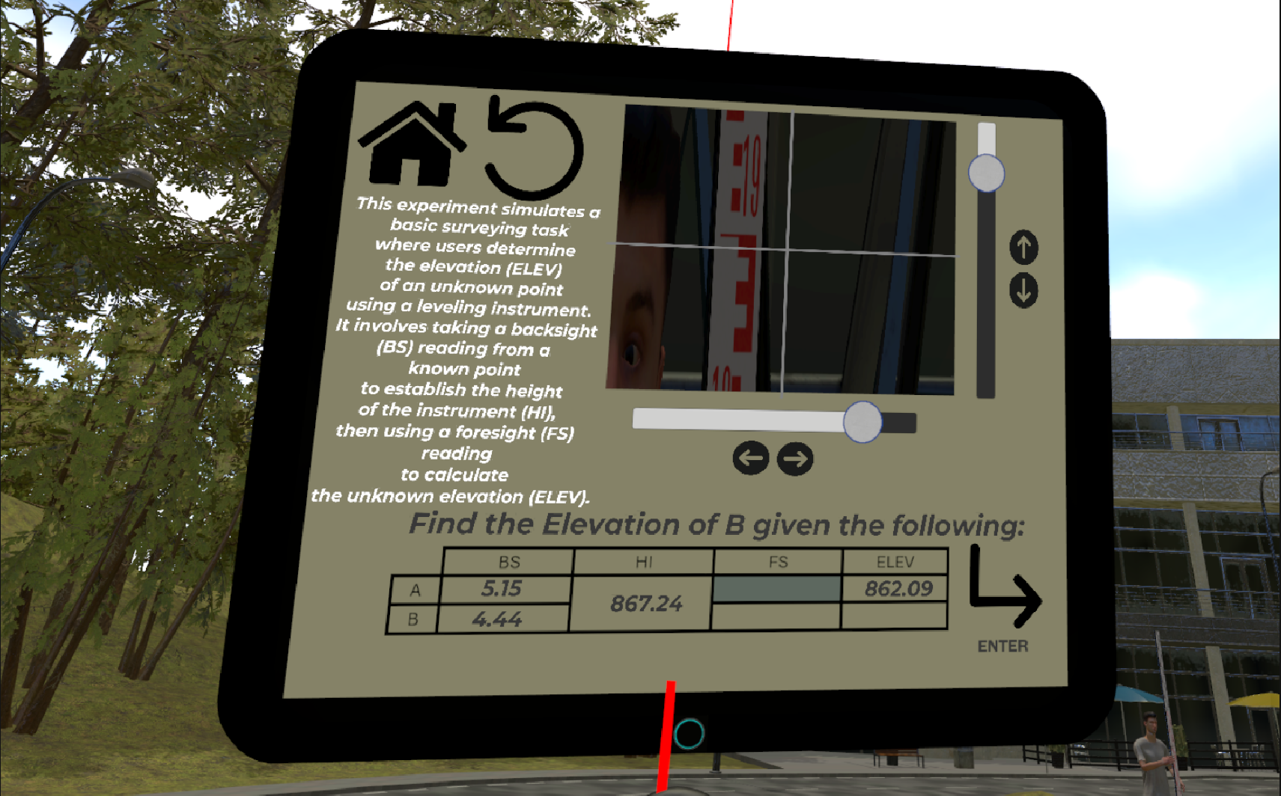}\label{fig:drone_control}}
\hfil
\subfloat[]{\includegraphics[width=0.49\textwidth]{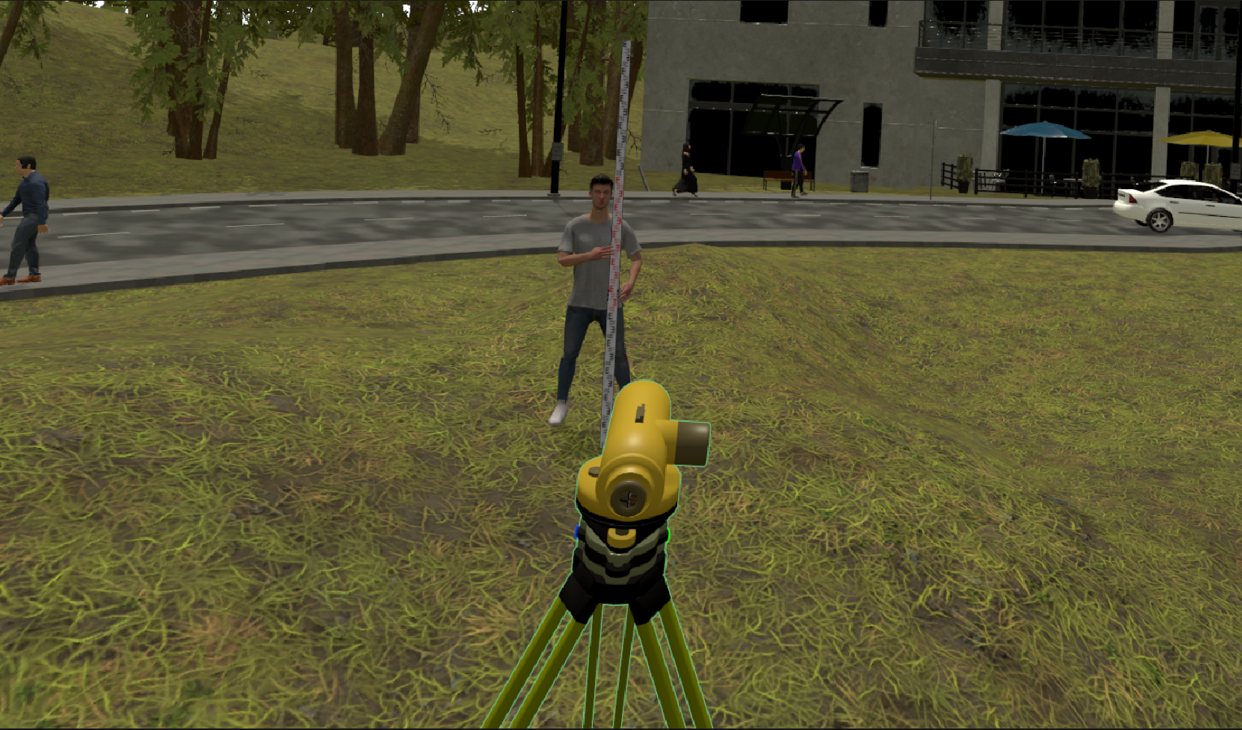}\label{fig:drone_flight}}
\caption{Computing elevation using differential leveling in VRISE: (a) Tablet interface during levelling; (b) Digital level aimed at leveling rod}
\label{fig:ground_survey_results}
\end{figure*}

The differential leveling task was designed to evaluate students' ability to apply fundamental surveying principles within a realistic VR context. As illustrated in Figure \ref{fig:ground_survey_results}, students were presented with a digital interface simulating a real-world problem: determining the elevation of an unknown point B using the known elevation at point A, a back sight (BS), and a foresight (FS).

\subsubsection{Positional Accuracy}
\begin{figure}[!h]
  \centering
  \includegraphics[width=0.5\textwidth]{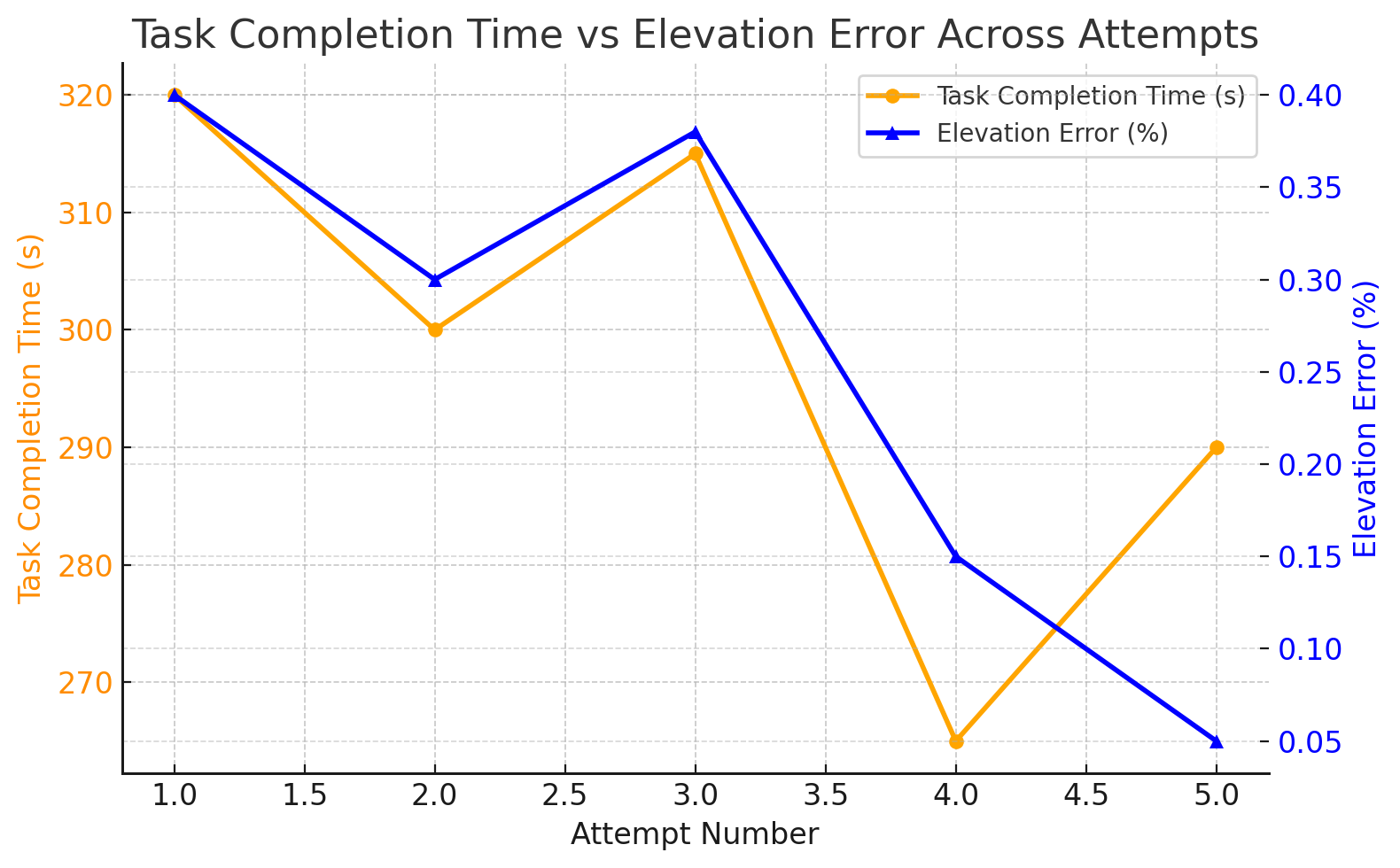}
      \caption{Positional accuracy improvement over five user attempts.} \label{fig:position}
\end{figure}

Figure \ref{fig:position} presents a line chart illustrating the elevation error (in \%) recorded over five consecutive attempts during the differential leveling exercise. Each attempt represents a full execution of the leveling task by a user, from instrument setup to final measurement. The use of five attempts enables observation of learning progression, allowing for a meaningful analysis of improvements in measurement accuracy over repeated practice. The chart reveals an overall improvement in precision, with error reducing from 0.4\% in Attempt 1 to just 0.05\% in Attempt 5. This downward trend is reflected in increased user proficiency in using leveling equipment and applying the correct techniques over time.
Notably, there is a slight anomaly in Attempt 3, where the error increased to 0.38\% after an earlier drop. This could indicate a momentary lapse in attention, incorrect instrument setup, or environmental interference. However, the sharp improvement in Attempts 4 and 5 suggests that the user was able to quickly correct the course and solidify their understanding of the process.
\newline
\subsubsection{Task Completion Time}
The general completion time trend in Figure \ref{fig:position} reflects performance improvement, with a noticeable reduction in time from the initial attempts. The first attempt took the longest (around 320 seconds), reflecting the user’s unfamiliarity with the procedure and possible hesitations or inefficiencies during setup and measurement phases.

By the fourth attempt, the completion time had dropped significantly to about 265 seconds,indicating a peak in efficiency and user adaptation. However, a slight increase in the fifth attempt suggests the possibility of a minor disruption, such as environmental distraction, fatigue, or a procedural error. Despite this, the overall downward trajectory in time supports the conclusion that repeated practice leads to improved procedural fluency and faster execution in leveling tasks.
\newline
\subsubsection{Interaction Efficiency}
\begin{figure}[!h]
  \centering
  \includegraphics[width=0.5\textwidth]{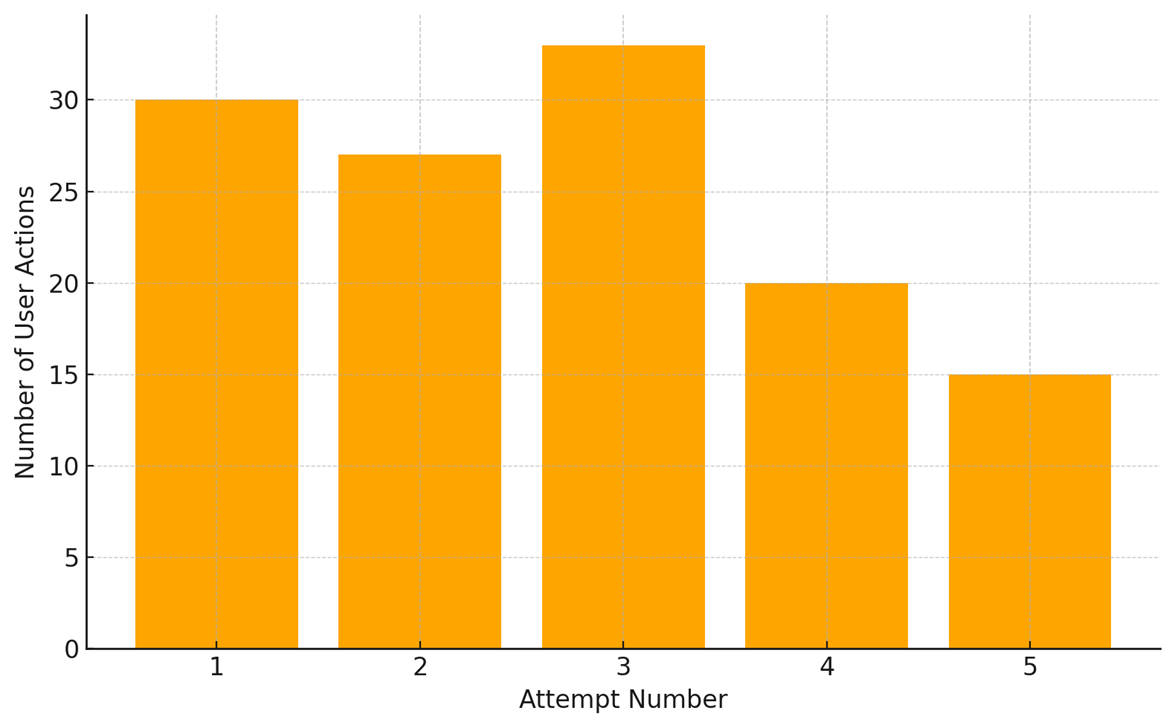}
      \caption{Number of user interactions per attempt.} \label{fig:interactionefficiency}
\end{figure}

The general trend shows a reduction in the number of user actions from Attempt 1 (30 actions) to attempt 5 (15 actions), reflecting a clear learning curve as shown in Figure \ref{fig:interactionefficiency}. This suggests that users become more familiar with the interface and the procedural steps of leveling over time, requiring fewer corrections, inputs, or confirmations to complete the task accurately.

Interestingly, attempt 3 stands out with a spike to 32 actions, indicating a momentary disruption in efficiency, possibly due to a mistake, recalibration, or confusion during the setup phase. Despite this anomaly, the subsequent decline in Attempts 4 and 5 suggests recovery and continued learning. Overall, the results demonstrate improving interaction efficiency, where the system supports user skill acquisition and minimizes unnecessary input as familiarity increases.

\subsection{AERIAL SURVEYING}
\begin{figure}[!h]
  \centering
  \includegraphics[width=0.5\textwidth]{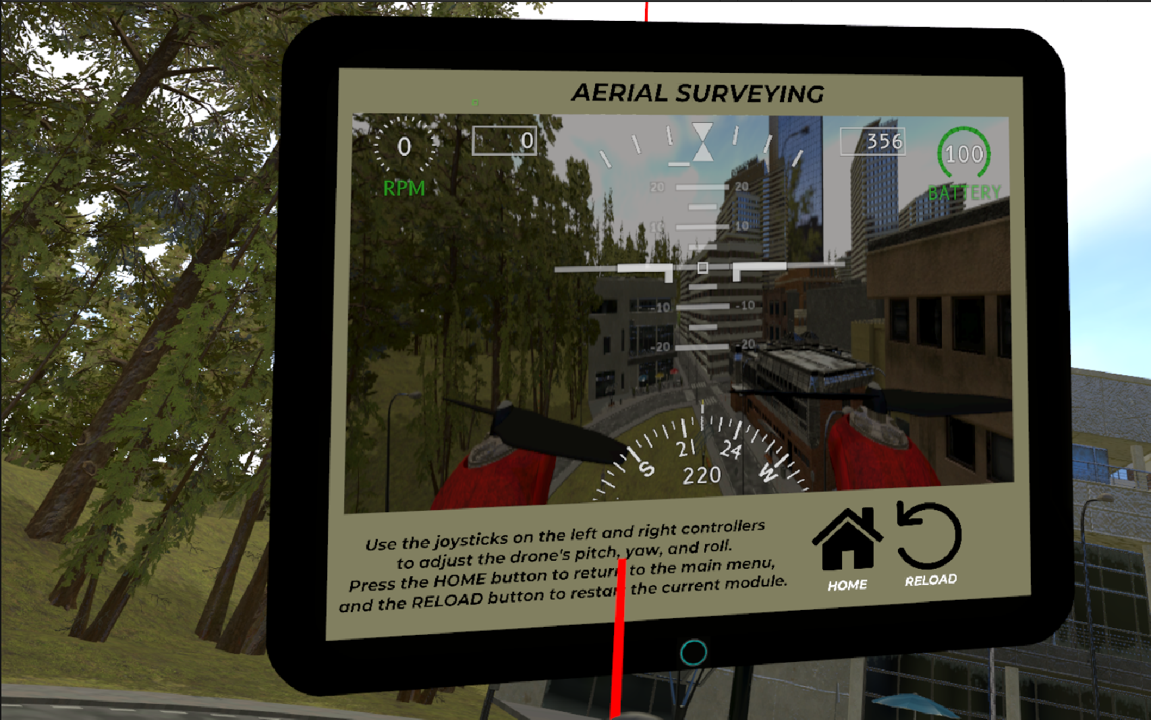}
      \caption{Tablet interface during the waypoint trailing exercise.} \label{fig:waypointresult}
\end{figure}

This section presents the results of the aerial surveying tasks conducted using the waypoint trailing exercise, as illustrated in the drone interface shown in Figure \ref{fig:waypointresult}. The image depicts the virtual control panel used in the simulation, where users piloted the drone through a predefined urban environment. The analysis focuses on the drone's ability to accurately follow designated paths, with key performance metrics, including waypoint completion accuracy, task completion time, and interaction efficiency, used to assess the overall effectiveness of the aerial surveying module.

\subsubsection{Waypoint Trailing Accuracy}
\begin{figure}[!h]
  \centering
  \includegraphics[width=0.5\textwidth]{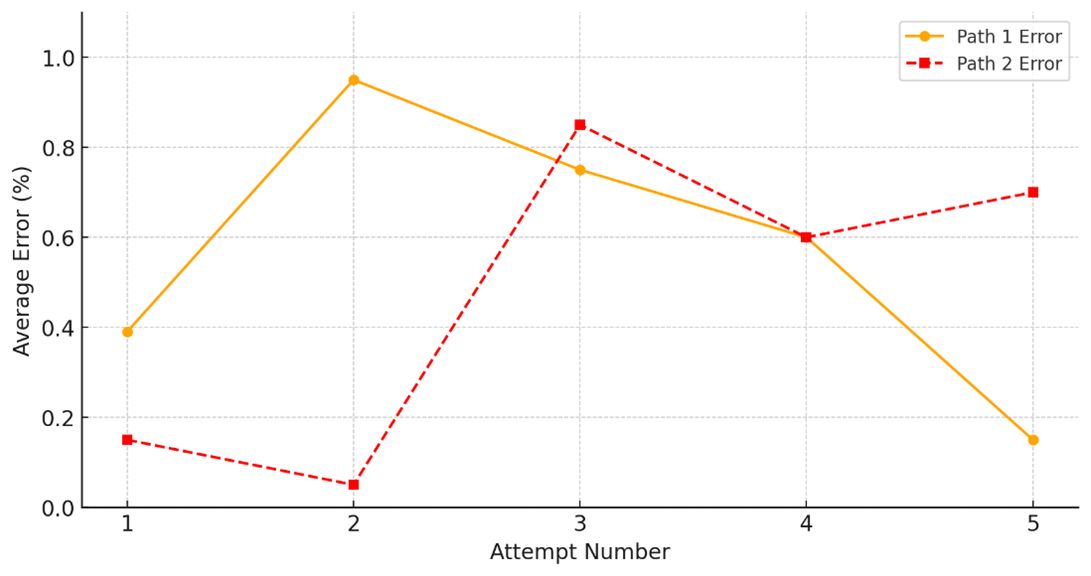}
      \caption{Trailing accuracy across the five attempts} \label{fig:waypointtraillingresult}
\end{figure}

Figure \ref{fig:waypointtraillingresult} presents the waypoint trailing accuracy, represented by the average error over five attempts for both Path 1 and Path 2. The error reflects the deviation between the user’s actual trajectory and the designated waypoints. For Path 1, the error reached its highest point during the second attempt likely due to early overcorrection or lack of familiarity but then showed a consistent decline, reaching its minimum by the fifth attempt. This downward trend suggests that users adapted quickly to the more direct path, progressively improving their positional accuracy with each attempt.

In contrast, Path 2 started with minimal error in the first two attempts, but experienced a noticeable spike in Attempt 3, likely due to the complexity of the curved path or overconfidence after early success. Although errors reduced again by Attempt 4, they remained consistently higher than Path 1 in later attempts, suggesting that Path 2 posed a more persistent challenge in maintaining accuracy. Overall, the data reflects that while both paths show signs of learning, Path 1 allows for quicker mastery, whereas Path 2 continues to demand higher precision and focus throughout.
\newline
\subsubsection{Task Completion Time}
\begin{figure}[H]
  \centering
  \includegraphics[width=0.5\textwidth]{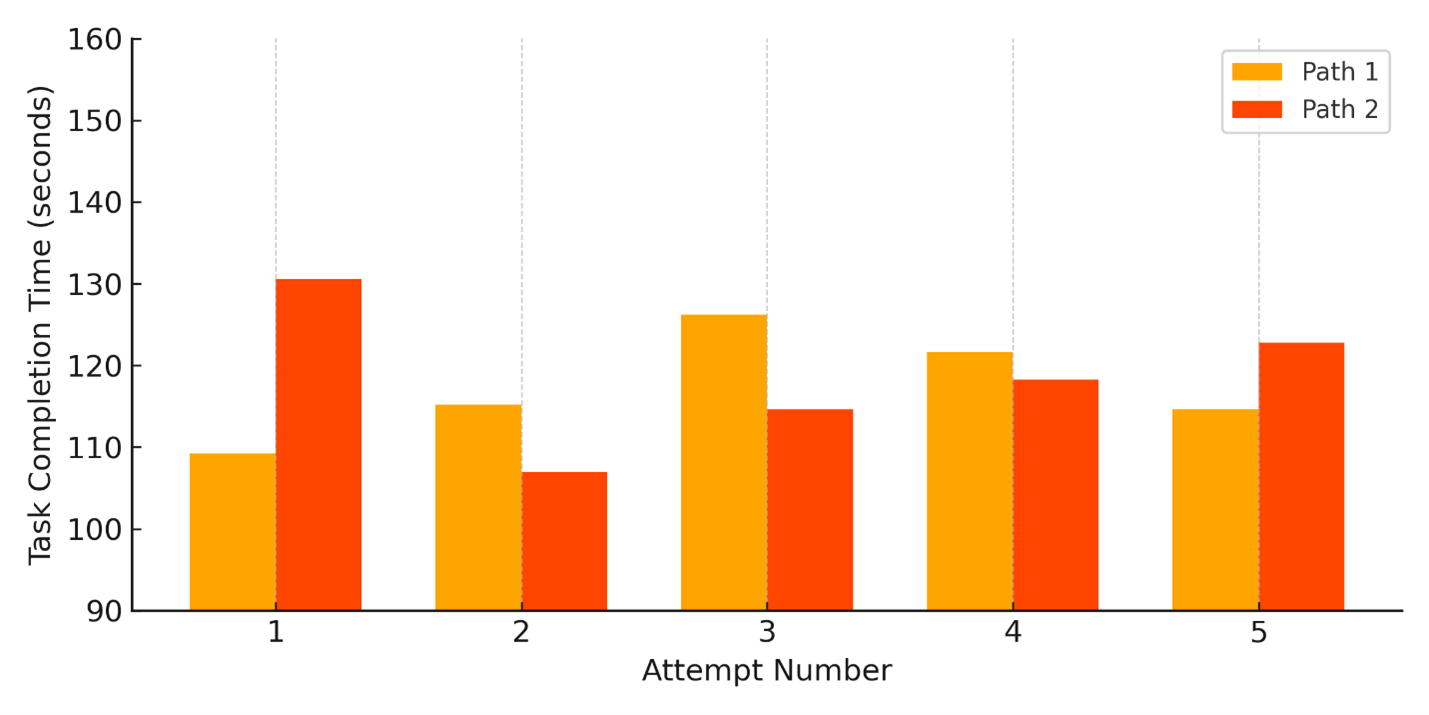}
      \caption{Task Completion times across the two paths of the waypoint trailing exercise} \label{fig:completiontime2}
\end{figure}

Figure \ref{fig:completiontime2} illustrates the task completion times for Path 1 and Path 2 over five attempts. Path 1 consistently maintained a relatively stable and efficient performance, with completion times ranging between approximately 110 and 127 seconds. This pattern suggests that users were able to quickly familiarize themselves with the short, direct nature of Path 1, achieving faster navigation and maintaining consistency throughout the trials.

Path 2, although longer and more curved, showed a more fluctuating performance. While the first attempt took notably longer (over 130 seconds), subsequent attempts reflect improvement, particularly in Attempt 2 where Path 2 was completed faster than Path 1. However, later attempts again required slightly more time, suggesting that while users may adapt to the complexity of Path 2, it still demands greater attention and precision, especially in the curved segments. Overall, the results indicate that Path 1 is more time-efficient, while Path 2’s performance varies based on user familiarity and control precision.
\newline
\subsubsection{Interaction Efficiency}
\begin{figure}[H]
  \centering
  \includegraphics[width=0.5\textwidth]{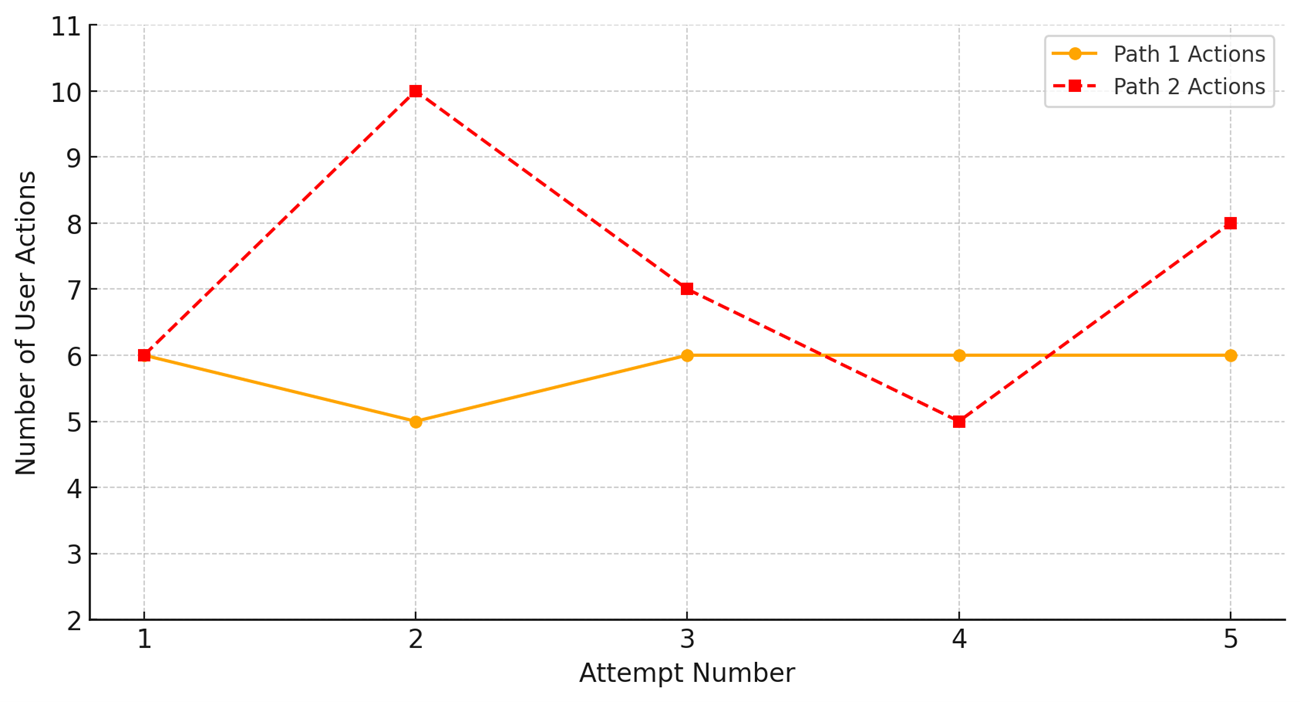}
      \caption{Task Completion times across the two paths of the waypoint trailing exercise} \label{fig:interactionefficiency2}
\end{figure}

Across five attempts, the number of user actions required showed that Path 1 consistently demanded fewer steps, typically ranging between 5 and 7 as shown in Figure \ref{fig:interactionefficiency2}. This narrow range indicates that users found the direct route easier to navigate, with minimal turns and corrections, resulting in smoother and more predictable performance. For expert users, completing Path 1 typically requires no more than two actions.

In contrast, Path 2 showed greater variability in the number of user actions, with a wider range resulting from its longer length and increased curvature. The higher frequency of interactions suggests users needed to make more frequent steering adjustments and corrections to complete the path. This indicates that Path 2 is more complex, requiring enhanced cognitive and motor control especially during initial attempts. However, for expert users, no more than five actions are typically needed to navigate Path 2 successfully.

\section{CONCLUSION AND LIMITATIONS} \label{conclusion}
This study introduced VRISE, a virtual civil engineering laboratory designed to provide accessible, immersive, and pedagogically effective surveying instruction for students. By integrating differential leveling and aerial surveying modules within a VR platform, VRISE offers a flexible and inclusive alternative to traditional field labs. The incorporation of assistive technologies, such as controller smoothing, interaction feedback, and customizable UI elements, enables learners to build spatial, procedural, and technical skills while minimizing physical strain and cognitive overload.

Quantitative results from user trials demonstrated that repeated exposure to VR-based surveying tasks led to improved accuracy, reduced task completion time, and enhanced interaction efficiency. The findings suggest that VRISE can support not only conceptual understanding but also skill acquisition in core surveying competencies, particularly for students who benefit from self-paced, visually guided, and adaptive learning environments.

However, several limitations must be acknowledged. First, the current evaluation involved a single user with intermediate VR skill, limiting generalizability across diverse student populations. Second, while the VR environment replicates core surveying workflows, it does not yet incorporate real-world data variability (e.g., sensor noise, weather effects) that may influence field conditions. Third, the system currently focuses on two modules, differential leveling and waypoint trailing, leaving room for expansion into other surveying techniques such as GNSS mapping or traverse computations.

Future work will focus on conducting broader usability studies involving diverse learner populations, including individuals with varying degrees of technical proficiency, learning differences, and physical abilities. These studies will aim to evaluate the adaptability of VRISE’s interface, the accessibility of its instructional design, and its effectiveness in real-world educational settings. By collaborating with institutions that serve underrepresented or non-traditional student groups, the research team intends to gather comprehensive feedback on how well VRISE accommodates a spectrum of learning needs, thereby enhancing its inclusivity and pedagogical robustness.

In parallel, development efforts will expand the VRISE platform to include a wider range of surveying instruments and technical modules. Planned additions include simulations for GNSS-based mapping, traverse computations, topographic profiling, and real-time data logging tools. These enhancements will allow the platform to more closely mirror the full scope of contemporary surveying practices, ensuring that students receive training aligned with industry standards. Incorporating such tools will not only diversify the skill sets that learners can acquire but also increase the platform’s relevance for professional certification programs and advanced coursework.

Finally, the VRISE team will work toward aligning the platform’s instructional modules with formal learning outcomes defined by engineering accreditation bodies such as the Accreditation Board for Engineering and Technology (ABET). This will involve mapping each VR activity to targeted competencies in spatial reasoning, data interpretation, procedural fluency, and ethical practice. The integration of learning analytics and adaptive assessment tools will support this effort by enabling instructors to track student performance in real-time and adjust instruction accordingly. Ultimately, VRISE aspires to be more than a simulation tool, it aims to serve as a scalable, equity-focused educational solution that bridges the gap between virtual practice and real-world engineering expertise.

\bibliographystyle{IEEEtran}
\bibliography{reference.bib}

\newpage

%\bf{If you include a photo:}\vspace{-33pt}
\begin{IEEEbiography}[{\includegraphics[width=1in,height=1.25in,clip,keepaspectratio]{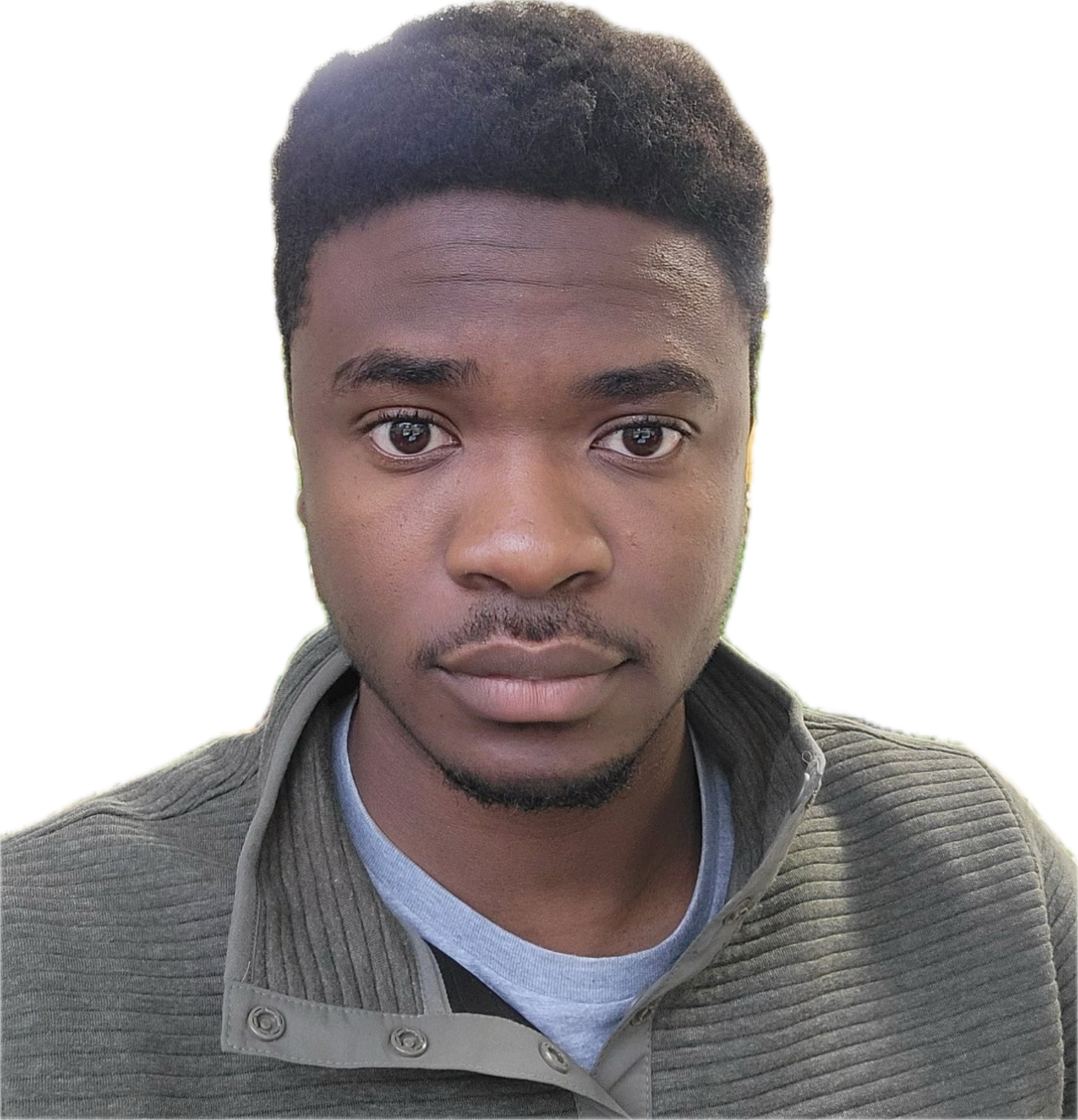}}]{Daniel Udekwe}
is a Ph.D. candidate in Civil Engineering at Florida State University, with research interests at the intersection of transportation resilience, intelligent systems, and immersive technologies. His work explores innovative methods to enhance data-driven decision-making in complex systems. Daniel is particularly interested in how emerging technologies such as virtual reality and quantum computing can be leveraged to create equitable, efficient, and scalable solutions across a range of civil engineering challenges.

\end{IEEEbiography}
\setlength{\parskip}{5pt}
\begin{IEEEbiography}[{\includegraphics[width=1in,height=1.25in,clip,keepaspectratio]{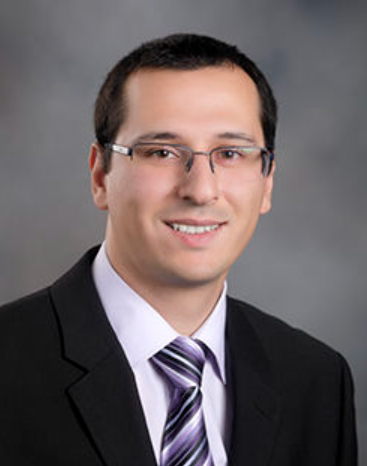}}]{Dimitrios Bolkas}  Associate Professor and Program Chair of the Surveying Engineering program at the Pennsylvania State University. BS and M.Eng. in Surveying Engineering, from the Aristotle University of Thessaloniki, Greece, and a PhD in Geological Sciences and Engineering from Queen’s University, Canada. He has a diverse geospatial background, which includes terrestrial, mobile, and airborne laser scanning, UASs, 3D modeling, monitoring and change estimation. Other research endeavors include immersive virtual reality to support engineering education. Dr. Bolkas serves on international organizations such as ISPRS and FIG. He is also an Associate Editor of the ASCE Journal of Surveying Engineering.
\end{IEEEbiography}

\setlength{\parskip}{5pt}
\begin{IEEEbiography}[{\includegraphics[width=1in,height=1.25in,clip,keepaspectratio]{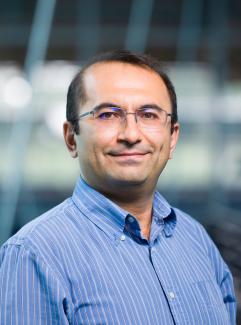}}]{Eren Erman Ozguven} is an Associate Professor in the Department of Civil and Environmental Engineering at the FAMU-FSU College of Engineering and Director of the Resilient Infrastructure and Disaster Response (RIDER) Center. Dr. Ozguven is focused on investigating the relationships among different infrastructure networks in Florida, utilizing his academic research in transportation engineering, and background in industrial engineering and optimization. His work examines the simultaneous and interdependent movements of populations—including the aging and other vulnerable groups, and the commodities and services that meet their needs. The research program he created draws from various engineering and science methods including optimization, statistical analysis, human factors, machine learning, traffic and transportation engineering and geography.
\end{IEEEbiography}

\setlength{\parskip}{5pt}
\begin{IEEEbiography}[{\includegraphics[width=1in,height=1.25in,clip,keepaspectratio]{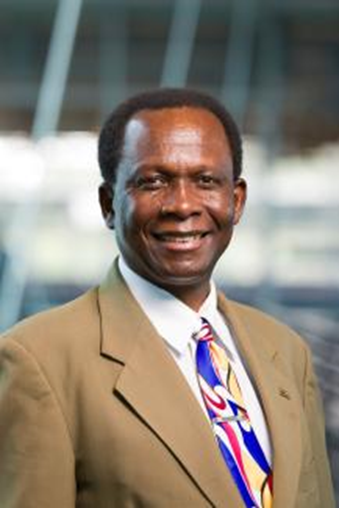}}]{Ren Moses} is a Professor in the Department of Civil and Environmental Engineering at the FAMU-FSU College of Engineering and Director of the Rural Equitable and Accessible Transportation (REAT) Center, a Tier 1 University Transportation Center funded by the U.S. Department of Transportation. His areas of expertise include traffic engineering and operations, highway safety, multimodal system design, and transportation workforce development.

\end{IEEEbiography}

\setlength{\parskip}{5pt}
\begin{IEEEbiography}[{\includegraphics[width=1in,height=1.25in,clip,keepaspectratio]{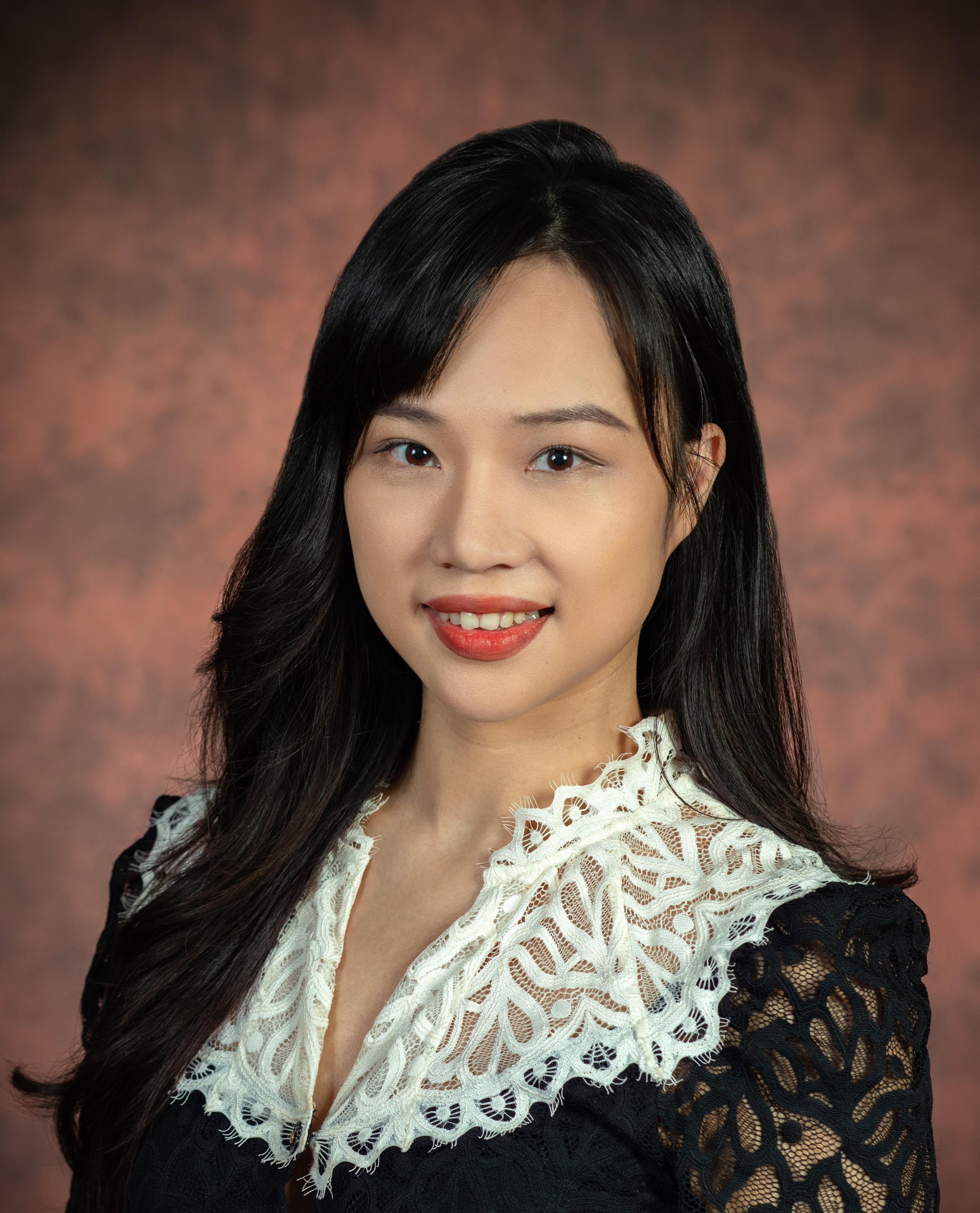}}]{Qian-wen (Vivian) Guo}
is an Assistant Professor of Transportation Engineering at Florida State University. She received her Ph.D. and M.S. in Management Science and Engineering from Huazhong University of Science and Technology and completed a joint Ph.D. program at Cornell University. Her research interest lies in Transportation Systems optimization problems, especially in public transit, ridesharing, phased development of civil infrastructure systems, as well as transportation economics. Dr. Guo is the principle investigator of multiple projects funded by National Science Foundation, US Department of Transportation, and Florida Department of Transportation.
\end{IEEEbiography}

\vspace{11pt}

\vfill

\end{document}